\documentclass[11pt]{article}

\usepackage[a4paper,margin=2.7cm]{geometry}
\usepackage{amsmath,amssymb,amsfonts}
\usepackage{graphicx}
\usepackage{xcolor}
\usepackage{colortbl}
\usepackage{array}
\usepackage{booktabs}
\usepackage{tabularx}
\usepackage{xltabular}
\usepackage{longtable}
\usepackage{caption}
\usepackage{subcaption}
\usepackage{float}
\usepackage{multicol}
\usepackage{authblk}
\usepackage[numbers,sort&compress]{natbib}
\usepackage[colorlinks=true,linkcolor=blue,citecolor=blue,urlcolor=blue]{hyperref}
\usepackage{orcidlink}

\graphicspath{{Figures/}}
\numberwithin{equation}{section}

\newcolumntype{C}{>{\centering\arraybackslash}X}
\newcolumntype{L}{>{\raggedright\arraybackslash}X}
\newcolumntype{R}{>{\raggedleft\arraybackslash}X}

\newcommand{\WGC}{WGC}
\newcommand{\WCCC}{WCCC}
\newcommand{\EMPYM}{EMPYM--AdS}
\newcommand{\Qcal}{\mathcal{Q}}
\newcommand{\Vol}{\mathcal{V}}

\title{\bfseries Holographic CFT Thermodynamics as a Bridge Between the Weak Gravity and Weak Cosmic Censorship Conjectures for EMPYM--AdS Black Holes}

\author[1]{Saeed Noori Gashti\,\orcidlink{0000-0001-7844-2640}}
\author[1,2]{Behnam Pourhassan\,\orcidlink{0000-0003-1338-7083}}
\author[3]{\.{I}zzet Sakall{\i}\,\orcidlink{0000-0001-7827-9476}\thanks{Corresponding author: izzet.sakalli@emu.edu.tr}}
\author[4]{Saheb Soroushfar\,\orcidlink{0000-0003-3151-0532}}

\affil[1]{School of Physics, Damghan University, P.~O.~Box 3671641167, Damghan, Iran}
\affil[2]{Center for Theoretical Physics, Khazar University, 41 Mehseti Street, Baku, AZ1096, Azerbaijan}
\affil[3]{Physics Department, Eastern Mediterranean University, Famagusta 99628, North Cyprus via Mersin 10, T\"urkiye}
\affil[4]{Department of Physics, College of Sciences, Yasouj University, 75918-74934 Yasouj, Iran}

\date{}

\begin{document}
\maketitle

\begin{abstract}
\noindent
The weak cosmic censorship conjecture (\WCCC) and the weak gravity conjecture (\WGC) sit at the heart of how singularities and charged matter are constrained in any consistent theory of quantum gravity. We study both conjectures for Einstein--Maxwell--power--Yang--Mills--AdS (\EMPYM) black holes, a family in which a non-Abelian power-law Yang--Mills field, a Maxwell field, and a negative cosmological constant act together. Working in holographic conformal field theory (CFT) thermodynamics, we treat the central charge $C$ and the CFT volume $\Vol$ as independent variables and obtain an extended first law that carries variations of the Yang--Mills charge and of the nonlinearity exponent $\gamma$. A closed Euler relation follows, and we check it together with the full set of equations of state to machine precision. The perturbation is a charged massive scalar minimally coupled to the Maxwell sector. Its horizon energy and charge fluxes fix the superradiance threshold $\omega<\tilde{q}_s\tilde{\phi}_h$, and the mass--energy relation $\mu_s=\omega$ turns this into the bound $\tilde{q}_s/\mu_s>r_{\min}/\tilde{Q}$. Expanding the metric function $f(r)$ about its minimum for extremal and near-extremal configurations, we find that absorption keeps $f_{\min}$ non-positive, so the horizon survives. The Yang--Mills sector lowers the effective charge-to-mass threshold below the Reissner--Nordstr\"om value of unity, while a local stability analysis locates a Davies point in the heat capacity. Across the parameter range examined the \EMPYM\ family respects cosmic censorship under scalar perturbations, with the \WGC\ fixing only the direction of evolution toward or away from extremality. We also extend the same analysis to the information content of these black holes through the island prescription. Working in that framework, we derive the Page curve and obtain an explicit Page time written in terms of the full set of thermodynamic variables. This ties the extended black hole thermodynamics, or black hole chemistry, directly to the recovery of information in Hawking radiation.
\end{abstract}

\section{Introduction}\label{isec1}

The Swampland Program~\cite{1} sorts effective field theories into those that admit a consistent ultraviolet completion with gravity and those that do not. The dividing surface, the swampland, is fixed by a set of conjectures that act as low-energy consistency criteria, and these criteria have grown into working tools for model building across particle physics and cosmology~\cite{2,3,13,22,26,28,32,38,39,43,57,60,62,63,64,66}. The \WGC\ is among the sharpest of them. It requires gravity to be the weakest force, which translates into the existence of a state whose charge-to-mass ratio exceeds that of an extremal black hole~\cite{4,5,14,23,70,1n}. That single demand reaches into the hierarchy problem, the spectrum of allowed particles, and the long-time fate of charged black holes~\cite{6,7,8,18,29}. A large body of work has since tested the \WGC\ and its relatives in inflation, in dark-energy model building, and in black-hole thermodynamics~\cite{9,10,11,12,15,16,19,20,21,24,25,27,30,31,33,34,35,36,37,40,41,44,45,46,48,49,51,52,54,55,56,59,61,65,67,8000,8001}.

Holographic CFT thermodynamics gives these questions a second face. When the central charge and the boundary volume are promoted to independent variables, the bulk first law maps onto a boundary description in which the internal energy of a dual CFT plays the central role~\cite{74,75,76,77,78,42,53,71}. The dictionary has been worked out for charged, rotating, and dyonic AdS black holes~\cite{79,80,81}, and McInnes argued that the \WGC\ itself has a clean holographic dual~\cite{McInnes:2020hdw}. Quyet recently showed, for Reissner--Nordstr\"om--AdS black holes dressed with perfect-fluid dark matter, that the \WGC\ favors a definite branch of the CFT thermodynamics~\cite{Quyet:2026haf}. These results make the boundary picture a natural place to ask whether two conjectures with very different origins can be made to coexist.

The non-Abelian sector deserves a word of its own. Power-law Yang--Mills fields generalize the standard quadratic action to $(F^{a}_{\mu\nu}F^{a\,\mu\nu})^{\gamma}$, and the exponent $\gamma$ controls how the gauge energy density falls off with radius. This single parameter reshapes the horizon structure, the temperature, and the extremal bound, which is why it offers a sharper test of the conjectures than a pure Maxwell charge. Black-string, cylindrical, and Chaplygin-type AdS solutions have shown how additional matter sectors change the thermodynamic phase structure and the scattering response~\cite{Ahmed:2025bstr,Ahmed:2025cyl,Debnath:2022chap}, and tunneling analyses in non-trivial backgrounds have probed the emission spectrum directly~\cite{Sakalli:2015gwh}. What has been missing is a treatment that carries a power-law Yang--Mills charge through the full holographic first law and then asks whether censorship survives. That is the gap this work addresses.

The \WCCC\ is the other half of the story. It states that the singularities produced in gravitational collapse stay hidden behind horizons, so that the exterior remains predictable~\cite{20,83,84,85,86,87}. Whether a test field can strip that horizon away has a long history, and the answer depends on how the field is set up. Extremal black holes resist destruction by test fields, as Nat\'ario, Queimada, and Vicente established~\cite{Natario:2016bay}. The picture is not uniformly protective, though. When a charged scalar is driven hard enough, explicit \WCCC\ violations appear in four-dimensional AdS~\cite{Crisford:2017gsv}, and the survival of the horizon has been tied directly to the \WGC~\cite{Horowitz:2019eum}. The same logic later became a source of constraints on low-energy effective theories~\cite{Chen:2020lbt}, while gedanken-experiment analyses sharpened the bookkeeping for Born--Infeld and dilaton-Lifshitz backgrounds~\cite{He:2019dra,Jiang:2020ald,58}. Extremal stability and the \WCCC\ in Kiselev spacetime have been examined as well~\cite{Anand:2025sek}.

A recurring theme is that extra matter can be the deciding factor. Adding quintessence, clouds of strings, or perfect-fluid dark matter can arrange for the \WGC\ and the \WCCC\ to hold at once~\cite{16,17,47,69,72,73,68,73n,82}. Anand and co-workers carried this through for charged AdS black holes in perfect-fluid dark matter using charged-scalar fluxes within CFT thermodynamics~\cite{Anand:2025wgc}, and Noori Gashti and Pourhassan reached the same kind of compatibility for the Kerr--Newman--Kiselev--Letelier black hole~\cite{Gashti:2025cqc}. Photon-sphere and lensing observables have meanwhile emerged as a route to confront these bounds with data, both for ModMax black holes~\cite{Gashti:2025modmax} and in gravity's rainbow~\cite{Gashti:2026lensing}. Related thermodynamic-stability and observational studies of charged and AdS backgrounds round out the picture~\cite{Sucu:2025crbh,Ahmed:2025bstr,Debnath:2022chap}, as do field-scattering analyses in non-trivial backgrounds~\cite{Ahmed:2025cyl,Sakalli:2015gwh}.

Most of this literature stops at the Maxwell sector. The role of a non-Abelian power-law gauge field, which carries its own charge and its own nonlinearity exponent, has been left largely open. We close part of that gap here. The object of study is the \EMPYM\ black hole~\cite{9999}, a four-dimensional solution carrying a Maxwell charge $Q$, a power-Yang--Mills field with charge parameter $q$ and exponent $\gamma$, and a negative cosmological constant. Because the Yang--Mills sector adds thermodynamic variables and reshapes the metric function, it lets us test both conjectures over a wider class of interactions than the Maxwell case allows. Our route is the following. We build the holographic dictionary and the extended first law, then derive a closed Euler relation and verify it numerically. The charged scalar field is added next, its horizon fluxes are computed, and the minimum of $f(r)$ is followed through absorption for extremal and near-extremal configurations. Local thermodynamic stability and the photon-sphere window round out the analysis, since both connect the model to observation.

The information paradox has sharpened our reading of semiclassical gravity. Work over the
past few years shows that the theory can be extended through the island prescription, which
adds a non-perturbative gravitational correction to the entropy of Hawking radiation. In this
construction the von Neumann entropy of the radiation is computed from the generalised
entropy functional, and the appearance of quantum extremal surfaces produces the Page
curve, removing the long-standing tension with unitary
evolution~\cite{Almheiri2021,Hashimoto2020,Yu2022,Lin2024}.

The prescription has been tested across many backgrounds. Schwarzschild, Reissner--Nordstr\"om,
and AdS black holes all reproduce the expected Page transition and the late-time saturation of
the entanglement entropy~\cite{Almheiri2021,Hashimoto2020,Lin2024}. Holographic formulations
add a further point: the Page time need not be read off only from the dynamics but can be
written in terms of the thermodynamic variables of the hole, which ties entanglement dynamics
to black hole chemistry~\cite{Kubiznak2012,Kubiznak2017,Gunasekaran2012}. That link is the one
we follow here.

Extended black hole thermodynamics treats the cosmological constant as a pressure and the
mass as an enthalpy. Read this way, the island construction inherits a dependence on the full
phase-space structure of the
solution~\cite{Kubiznak2012,Gunasekaran2012,Kubiznak2017,Cai2013}, so information recovery is
not a process that stands apart from thermodynamic criticality but one that can carry the
signatures of phase transitions and of the local stability of the hole. The two are tied. That
is the thread we follow.

These observations motivate the present extension. We bring the island prescription to
Einstein--Maxwell--power--Yang--Mills--AdS (EMPYM--AdS) black
holes~\cite{Mazharimousavi2009}, a setting that suits the question because it carries both a
Maxwell sector and a nonlinear Yang--Mills sector tuned by the exponent $\gamma$, and the two
together reshape the horizon structure and the thermodynamic response. Nonlinear gauge fields
are known to move the stability boundaries, the phase structure, and the scattering response of
AdS black holes~\cite{Wei2013,Cai2013,Mazharimousavi2009}. What is new here is the explicit
Page time we obtain for this class, written in terms of the Maxwell charge, the Yang--Mills
parameter, the nonlinear exponent, and the pressure, so that the machinery of extended black
hole chemistry can be read directly as a control on information recovery.

The article is organized as follows. Section~\ref{isec2} sets up the \EMPYM\ solution, the holographic dictionary, the equations of state, the Euler relation, and the heat-capacity analysis. In Section~\ref{isec3} we add the charged scalar field and compute the horizon fluxes. The censorship test under absorption, treated in Section~\ref{isec4}, covers both the extremal and the near-extremal cases. We then turn to mass--energy equivalence in Section~\ref{isec5}, where the \WGC\ threshold is extracted and tied to the photon sphere. A parameter survey with cross-checks fills Section~\ref{isec6}. We then turn to the island
formula and information recovery (Section~\ref{sec:island}), where the Page curve and the Page
time are worked out for the present model. Section~\ref{isec7} gathers the conclusions. Throughout we work in geometric units $G=c=\hbar=1$ and in four spacetime dimensions, $n=2$.

\section{The EMPYM--AdS black hole and its holographic CFT thermodynamics}\label{isec2}

This section fixes the bulk solution and carries it to the boundary. We begin with the static
\EMPYM\ metric and its horizon structure, read off the bulk thermodynamics, and set up the
holographic dictionary in which the central charge $C$ and the CFT volume $\Vol$ are treated as
independent variables. The boundary equations of state follow, together with a closed Euler
relation that we verify numerically, the local stability read from the heat capacity, and the
on-shell free energy. Each piece feeds the censorship and weak-gravity analysis taken up in the
later sections.

\subsection{Bulk solution and horizon structure}\label{isec2a}

We take a static, spherically symmetric \EMPYM\ black hole with line element~\cite{9999}
\begin{equation}
ds^{2}=-f(r)\,dt^{2}+\frac{dr^{2}}{f(r)}+r^{2}\,d\Omega_{n}^{2},
\label{eq:metric}
\end{equation}
where $d\Omega_{n}^{2}$ is the metric on a unit $n$-sphere with $n\ge2$, and we work in the four-dimensional case $n=2$. The cosmological constant is $\Lambda=-n(n+1)/(2\ell^{2})$, with $\ell$ the AdS curvature radius. The metric function reads
\begin{equation}
f(r)=1-\frac{2GM}{r^{\,n-1}}+\frac{n(n+1)}{6}\frac{r^{2}}{\ell^{2}}+\frac{2(n-1)Q^{2}}{n\,r^{2n-2}}-\frac{\Qcal}{r^{4\gamma-2}},
\label{eq:f}
\end{equation}
where $M$ is the ADM mass, $Q$ the Maxwell charge, $G$ Newton's constant, and $\gamma$ the nonlinearity exponent of the power-Yang--Mills field. The Yang--Mills amplitude is
\begin{equation}
\Qcal=\frac{G\big[(n-1)n\,q^{2}\big]^{\gamma}}{n(4\gamma-n-1)},
\label{eq:Qcal}
\end{equation}
with $q$ the Yang--Mills charge parameter. The power-Yang--Mills Lagrangian density is proportional to $(F^{a}_{\mu\nu}F^{a\,\mu\nu})^{\gamma}$, where $F^{a}_{\mu\nu}$ is the non-Abelian field strength. The construction returns the standard Yang--Mills theory at $\gamma=1$, and the exponent $\gamma$ introduces a new scale that reshapes both the thermodynamics and the response to external perturbations. For $n=2$ and $G=1$ the amplitude simplifies to $\Qcal=(2q^{2})^{\gamma}/[2(4\gamma-3)]$, which requires $\gamma>3/4$ for a finite positive value.

\begin{figure}[htbp]
  \centering
  \begin{subfigure}[t]{0.49\textwidth}
    \centering
    \includegraphics[width=\textwidth]{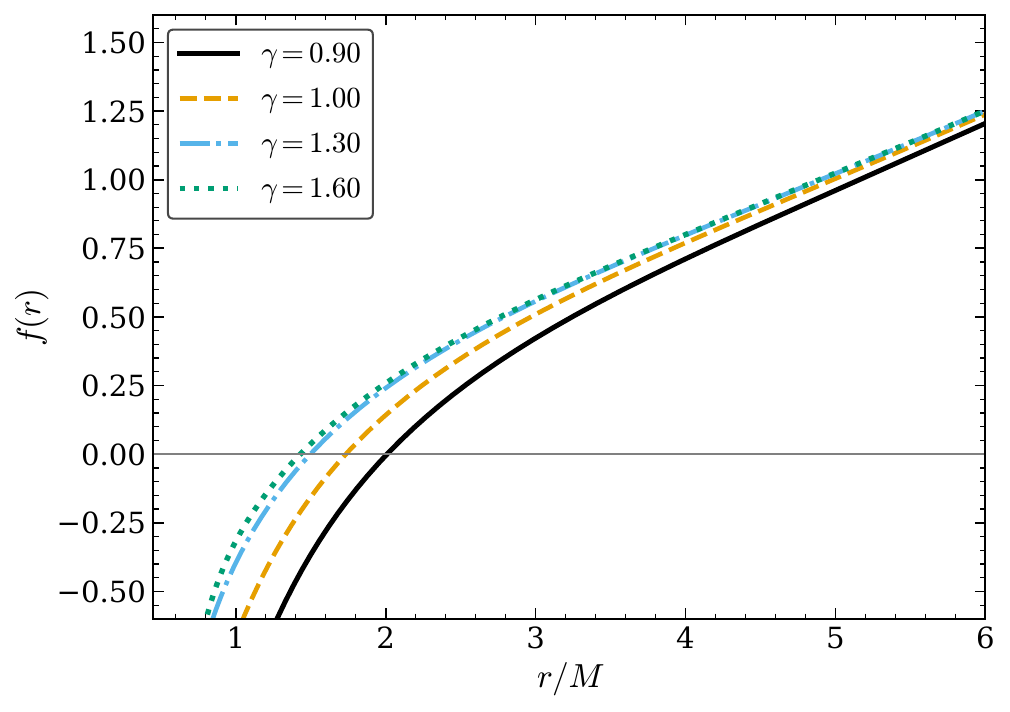}
    \caption{Varying the exponent $\gamma=0.90,1.00,1.30,1.60$ at $M=1$, $Q=0.9$, $q=0.7$, $\ell=8$.}
    \label{fig:f_gamma}
  \end{subfigure}\hfill
  \begin{subfigure}[t]{0.49\textwidth}
    \centering
    \includegraphics[width=\textwidth]{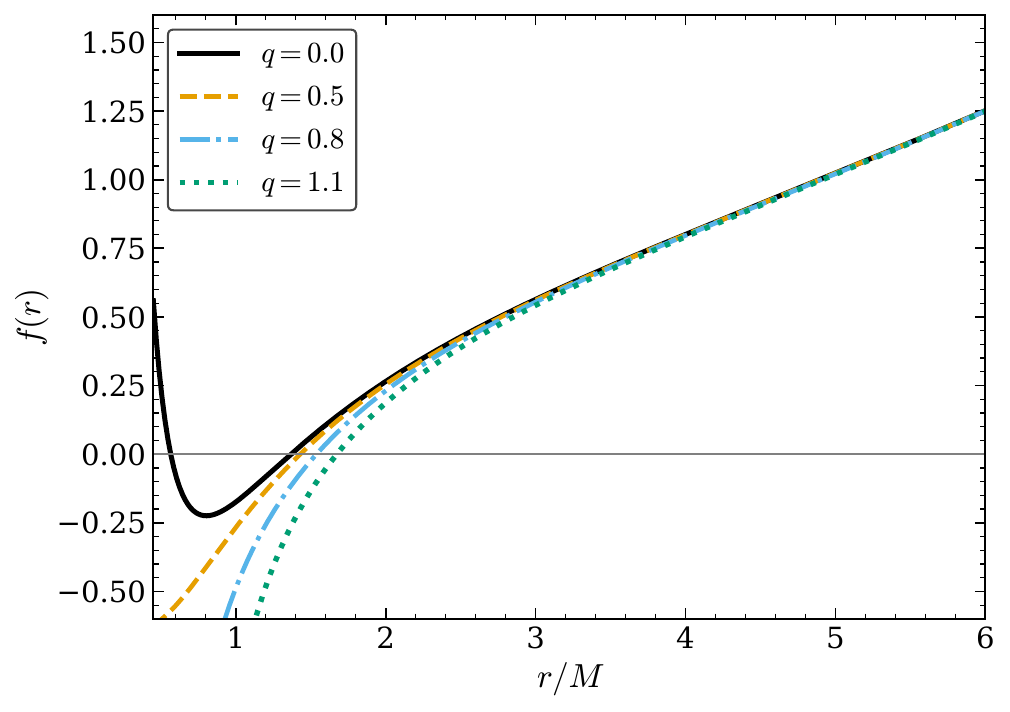}
    \caption{Varying the Yang--Mills charge $q=0.0,0.5,0.8,1.1$ at $\gamma=1.3$, $M=1$, $Q=0.9$, $\ell=8$; the $q=0$ curve is the Reissner--Nordstr\"om--AdS reference.}
    \label{fig:f_q}
  \end{subfigure}
  \caption{Metric function $f(r)$ versus $r/M$ for the \EMPYM\ black hole.}
  \label{fig:f_both}
\end{figure}

The horizon structure follows from the roots of $f(r)=0$. Figure~\ref{fig:f_gamma} plots $f(r)$ for four values of $\gamma$ at fixed $M$, $Q$, $q$, and $\ell$. The outer root moves inward as $\gamma$ grows, since a larger exponent suppresses the Yang--Mills amplitude $\Qcal$ through the factor $(2q^{2})^{\gamma}/(4\gamma-3)$ and shrinks its reach in $r$. The same trend, read against the Maxwell charge, appears in Fig.~\ref{fig:f_q}.

Why does the outer horizon move at all? Equation~\eqref{eq:Qcal} answers it: as $\Qcal$ falls with increasing $\gamma$, the negative Yang--Mills term in Eq.~\eqref{eq:f} weakens and the metric crosses zero at a smaller radius. For the two smallest exponents the curve keeps a positive Maxwell-dominated branch at small $r$, so an inner horizon survives; once $\gamma\gtrsim1.3$ the Yang--Mills term governs the small-$r$ behavior and only a single horizon is left. This is the geometric counterpart of the change in $\Qcal$ recorded in Table~\ref{tab:horizon}, and in the $q\to0$ limit, where the Yang--Mills term drops out, it reduces to the Reissner--Nordstr\"om--AdS pattern.

The dependence on the Yang--Mills charge in Fig.~\ref{fig:f_q} runs the other way. Raising $q$ enlarges $\Qcal$ through the factor $(2q^{2})^{\gamma}$, which deepens the negative term in Eq.~\eqref{eq:f} and pushes the outer horizon outward. The $q=0$ curve coincides with the Reissner--Nordstr\"om--AdS solution, so the spread between curves measures the cumulative effect of the non-Abelian sector.

\begin{table*}[htbp]
  \centering
  \setlength{\tabcolsep}{7pt}
  \renewcommand{\arraystretch}{1.25}
  \begin{tabularx}{\textwidth}{C C C C C C}
    \toprule
    \rowcolor{orange!50}
    $\gamma$ & $\Qcal$ & $r_{\min}$ & $f_{\min}$ & $r_{h}$ (outer) & horizons \\
    \midrule
    0.90 & 0.81832 & 0.36991 & $-2.50260$ & 2.00860 & 2 \\
    1.00 & 0.49000 & 0.31984 & $-2.12340$ & 1.73400 & 2 \\
    1.30 & 0.22138 & --      & --         & 1.49770 & 1 \\
    1.60 & 0.14238 & --      & --         & 1.43010 & 1 \\
    2.00 & 0.09604 & --      & --         & 1.39520 & 1 \\
    \bottomrule
  \end{tabularx}
  \caption{Horizon structure of the \EMPYM\ black hole as $\gamma$ varies, at $M=1$, $Q=0.9$, $q=0.7$, $\ell=8$ (geometric units). Columns list the Yang--Mills amplitude $\Qcal$ from Eq.~\eqref{eq:Qcal}, the location and value of the interior minimum of $f$ where it exists, the outer horizon radius, and the horizon count. A dash marks configurations with no interior minimum outside the horizon.}
  \label{tab:horizon}
\end{table*}

Across the same range of $\gamma$, Table~\ref{tab:horizon} records how the horizon content changes. The amplitude $\Qcal$ drops by almost an order of magnitude between $\gamma=0.9$ and $\gamma=2.0$, and near $\gamma\simeq1.3$ the system passes from two horizons to one. The reason sits in the small-$r$ limit of Eq.~\eqref{eq:f}, where the Maxwell term $Q^{2}/r^{2}$ competes with the Yang--Mills term $\Qcal\,r^{2-4\gamma}$: once $4\gamma-2>2$, the latter wins and removes the inner horizon. The change is smooth, and the outer horizon stays intact, so a regular exterior persists throughout.

\subsection{Bulk thermodynamics and the holographic map}\label{isec2bulk}

Before turning to the boundary description it helps to fix the bulk thermodynamics, since the holographic dictionary maps these quantities to the CFT. The horizon $r_{h}$ is the largest root of $f(r_{h})=0$, and the Hawking temperature follows from the surface gravity,
\begin{equation}
T=\frac{f'(r_{h})}{4\pi}=\frac{1}{4\pi}\left[\frac{2M}{r_{h}^{2}}+\frac{2r_{h}}{\ell^{2}}-\frac{2Q^{2}}{r_{h}^{3}}+\frac{(4\gamma-2)\Qcal}{r_{h}^{4\gamma-1}}\right].
\label{eq:THawking}
\end{equation}
The Bekenstein--Hawking entropy is one quarter of the horizon area, which for $n=2$ gives
\begin{equation}
S=\frac{A}{4G}=\pi r_{h}^{2},
\label{eq:Sentropy}
\end{equation}
so that $r_{h}=\sqrt{S/\pi}$. The bulk first law reads $dM=T\,dS+\phi\,dQ+\psi\,dq$, with $\phi=Q/r_{h}$ the electric potential and $\psi$ the Yang--Mills potential conjugate to $q$. Eliminating $M$ between $f(r_{h})=0$ and Eq.~\eqref{eq:THawking} expresses the mass as a function of the horizon radius,
\begin{equation}
M=\frac{r_{h}}{2}\left(1+\frac{r_{h}^{2}}{\ell^{2}}+\frac{Q^{2}}{r_{h}^{2}}-\Qcal\,r_{h}^{2-4\gamma}\right),
\label{eq:Mbulk}
\end{equation}
which is the ADM mass at the horizon. The temperature~\eqref{eq:THawking} vanishes at the extremal radius, where $f$ and $f'$ share a root, and this is the bulk image of the boundary extremal state where $\tilde{T}=0$ in Fig.~\ref{fig:T_S}. The map $E=M/\omega$, $\tilde{T}=T/\omega$, $\tilde{S}=S$ then carries Eqs.~\eqref{eq:THawking}--\eqref{eq:Mbulk} to the boundary, where the central charge and the volume become the new independent variables introduced below. Because the entropy~\eqref{eq:Sentropy} is shared by bulk and boundary, $\tilde{S}=S=\pi r_{h}^{2}$, the censorship analysis of Sec.~\ref{isec4}, which works with $f(r)$ directly, and the thermodynamic analysis of the boundary CFT refer to the same horizon data.

\subsection{Holographic dictionary and the extended first law}\label{isec2b}

To study the \WCCC\ and the \WGC\ we use holographic CFT thermodynamics. The central charge of the dual CFT is fixed by the AdS radius and Newton's constant, $C=\ell^{2}/(4G)$ for $n=2$, with the unit two-sphere volume $\Omega_{2}=4\pi$. We set $G=1$ from here on. The boundary CFT lives on a sphere of radius $R$, and a dimensionless conformal factor $\omega=R/\ell$ relates bulk and boundary quantities. The CFT spatial volume is $\Vol=4\pi R^{2}$, and the boundary pressure satisfies the conformal equation of state $p=E/(2\Vol)$, with $E$ the CFT internal energy. The dictionary connecting bulk and boundary variables is
\begin{equation}
E=\frac{M}{\omega},\quad \tilde{S}=S,\quad \tilde{T}=\frac{T}{\omega},\quad \tilde{Q}=\frac{Q\ell}{\sqrt{G}}=2Q\sqrt{C},\quad \tilde{\phi}=\frac{\phi}{R},\quad \tilde{q}=\frac{q\ell}{\sqrt{G}}=2q\sqrt{C},\quad \tilde{\psi}=\frac{\psi}{R},
\label{eq:dict}
\end{equation}
where $S$ is the entropy, $T$ the Hawking temperature, $\phi=Q/r_{h}$ the horizon electric potential, $\psi$ the Yang--Mills potential, and tilde quantities denote boundary counterparts. The chemical potential conjugate to $C$ is
\begin{equation}
\mu=\frac{1}{C}\left(E-\tilde{T}\tilde{S}-\tilde{\phi}\tilde{Q}-\frac{3\gamma-1}{2\gamma}\tilde{\psi}\tilde{q}\right).
\label{eq:mu_def}
\end{equation}
Using the extended thermodynamic relations and the definition of the CFT volume, the internal energy of the boundary CFT for the \EMPYM\ black hole is
\begin{equation}
E=\frac{2^{-\gamma-1}\pi^{2\gamma-\frac{3}{2}}C^{\frac{1}{2}-\gamma}S^{\frac{3}{2}-2\gamma}\tilde{q}^{2\gamma}}{(4\gamma-3)R}+\frac{\sqrt{\pi}\,\tilde{Q}^{2}}{4\sqrt{C}\,R\sqrt{S}}+\frac{S^{3/2}}{4\pi^{3/2}\sqrt{C}\,R}+\frac{\sqrt{C}\sqrt{S}}{\sqrt{\pi}\,R}.
\label{eq:E}
\end{equation}
The conformal factor obeys $\omega=R/\ell$, whose logarithmic variation gives
\begin{equation}
\frac{d\omega}{\omega}=\frac{dR}{R}-\frac{d\ell}{\ell},\qquad \frac{d\ell}{\ell}=\frac{1}{2}\frac{d\ell^{2}}{\ell^{2}},\qquad \frac{dR}{R}=\frac{1}{2}\frac{dR^{2}}{R^{2}}.
\label{eq:dom}
\end{equation}
Combining these relations with the bulk first law, the dictionary, and $C=\ell^{2}/(4G)$, the first law in boundary variables becomes
\begin{equation}
d\!\left(\frac{M}{\omega}\right)=\frac{T}{\omega}dS+\frac{\phi}{R}d(Q\ell)+\frac{\psi}{R}d(q\ell)+\left[\frac{M}{\omega}-\frac{T}{\omega}-\frac{\phi}{R}(Q\ell)-\frac{3\gamma-1}{2\gamma}\frac{\psi}{R}(q\ell)\right]\frac{dC}{C}-\frac{M}{2\omega}\frac{d\Vol}{\Vol}.
\label{eq:firstlaw_long}
\end{equation}
Applying the rescalings in Eq.~\eqref{eq:dict}, the first law takes the compact holographic form
\begin{equation}
dE=\tilde{T}\,dS+\tilde{\phi}\,d\tilde{Q}+\tilde{\psi}\,d\tilde{q}+\mu\,dC-p\,d\Vol,
\label{eq:firstlaw}
\end{equation}
with $\mu$ conjugate to $C$ and $p$ the boundary pressure on $\Vol=4\pi R^{2}$. Explicitly,
\begin{equation}
\mu=\frac{1}{C}\left(E-\tilde{T}S-\tilde{\phi}\tilde{Q}-\frac{3\gamma-1}{2\gamma}\tilde{\psi}\tilde{q}\right),\qquad p=\frac{E}{2\Vol}.
\label{eq:mu_p}
\end{equation}

\subsection{CFT equations of state}\label{isec2c}

All boundary thermodynamic quantities follow from $E$. The temperature is
\begin{equation}
\tilde{T}=\left(\frac{\partial E}{\partial S}\right)_{\tilde{Q},\tilde{q},C,\Vol}=-\frac{2^{-\gamma-2}\pi^{2\gamma-\frac{3}{2}}C^{\frac{1}{2}-\gamma}S^{\frac{1}{2}-2\gamma}\tilde{q}^{2\gamma}}{R}-\frac{\sqrt{\pi}\,\tilde{Q}^{2}}{8\sqrt{C}\,R\,S^{3/2}}+\frac{3\sqrt{S}}{8\pi^{3/2}\sqrt{C}\,R}+\frac{\sqrt{C}}{2\sqrt{\pi}\,R\sqrt{S}}.
\label{eq:T}
\end{equation}
The electric potential conjugate to $\tilde{Q}$ is
\begin{equation}
\tilde{\phi}=\left(\frac{\partial E}{\partial \tilde{Q}}\right)_{S,\tilde{q},C,\Vol}=\sqrt{\frac{\pi}{CS}}\,\frac{\tilde{Q}}{2R}.
\label{eq:phi}
\end{equation}
The Yang--Mills potential conjugate to $\tilde{q}$ is
\begin{equation}
\tilde{\psi}=\left(\frac{\partial E}{\partial \tilde{q}}\right)_{S,\tilde{Q},C,\Vol}=\frac{2^{-\gamma}\pi^{2\gamma-\frac{3}{2}}\gamma\,S^{\frac{3}{2}-2\gamma}}{(4\gamma-3)R}\,C^{\frac{1}{2}-\gamma}\tilde{q}^{2\gamma-1}.
\label{eq:psi}
\end{equation}
The chemical potential associated with the central charge is
\begin{equation}
\mu=\left(\frac{\partial E}{\partial C}\right)_{S,\tilde{Q},\tilde{q},\Vol}=\frac{\sqrt{S}}{2\sqrt{\pi C}\,R}-\frac{S^{3/2}}{8\pi^{3/2}C^{3/2}R}+\frac{2^{-\gamma-1}\pi^{2\gamma-\frac{3}{2}}\big(\tfrac{1}{2}-\gamma\big)C^{-\gamma-\frac{1}{2}}S^{\frac{3}{2}-2\gamma}\tilde{q}^{2\gamma}}{(4\gamma-3)R}-\frac{\sqrt{\pi}\,\tilde{Q}^{2}}{8C^{3/2}R\sqrt{S}}.
\label{eq:mu}
\end{equation}
Finally, the boundary pressure follows from the volume derivative,
\begin{equation}
p=\left(\frac{\partial E}{\partial \Vol}\right)_{S,\tilde{Q},\tilde{q},C}=\frac{1}{8\pi R^{2}}\left(\frac{2^{-\gamma-1}\pi^{2\gamma-\frac{3}{2}}C^{\frac{1}{2}-\gamma}S^{\frac{3}{2}-2\gamma}\tilde{q}^{2\gamma}}{(4\gamma-3)R}+\frac{\sqrt{\pi}\,\tilde{Q}^{2}}{4\sqrt{C}\,R\sqrt{S}}+\frac{S^{3/2}}{4\pi^{3/2}\sqrt{C}\,R}+\frac{\sqrt{C}\sqrt{S}}{\sqrt{\pi}\,R}\right).
\label{eq:p}
\end{equation}
These expressions form the complete set of equations of state for the boundary CFT dual to the \EMPYM\ black hole. They carry the Maxwell charge $\tilde{Q}$, the Yang--Mills charge $\tilde{q}$, the exponent $\gamma$, and the central charge $C$, and they obey the extended first law~\eqref{eq:firstlaw}.

\begin{figure}[htbp]
  \centering
  \includegraphics[width=0.52\textwidth]{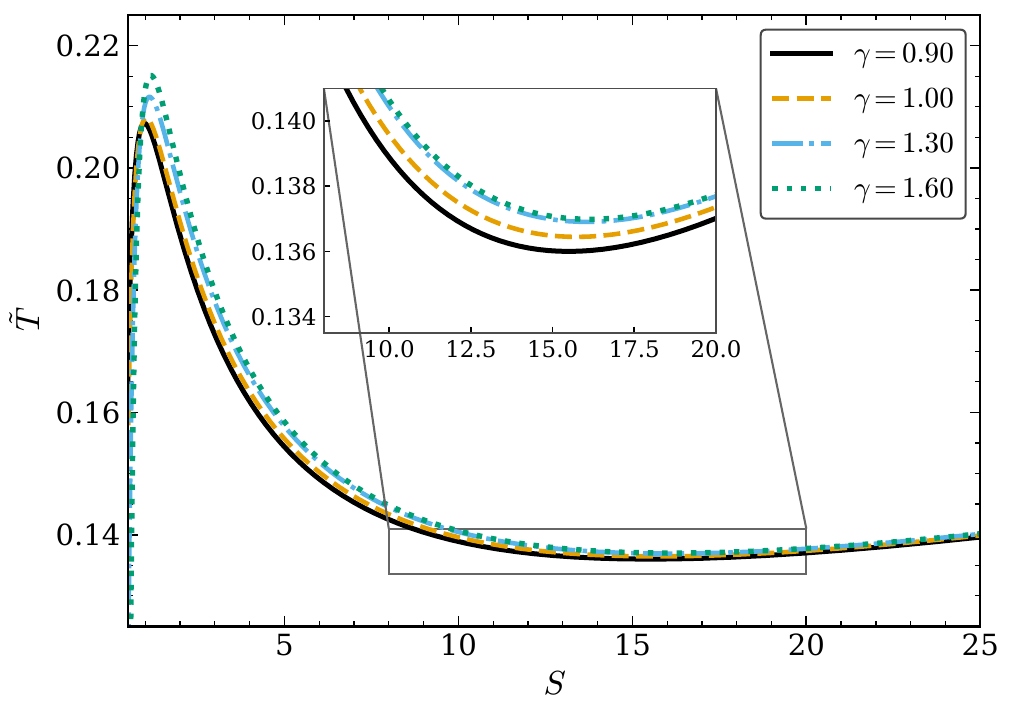}
  \caption{CFT temperature $\tilde{T}$ versus entropy $S$ for $\gamma=0.90,1.00,1.30,1.60$ at $C=4$, $\tilde{Q}=1$, $\tilde{q}=0.8$, $R=2$, from Eq.~\eqref{eq:T}. The inset magnifies the minimum region $S\in[8,20]$, where the four curves remain ordered but separate by less than a line width.}
  \label{fig:T_S}
\end{figure}

\begin{figure}[htbp]
  \centering
  \includegraphics[width=0.52\textwidth]{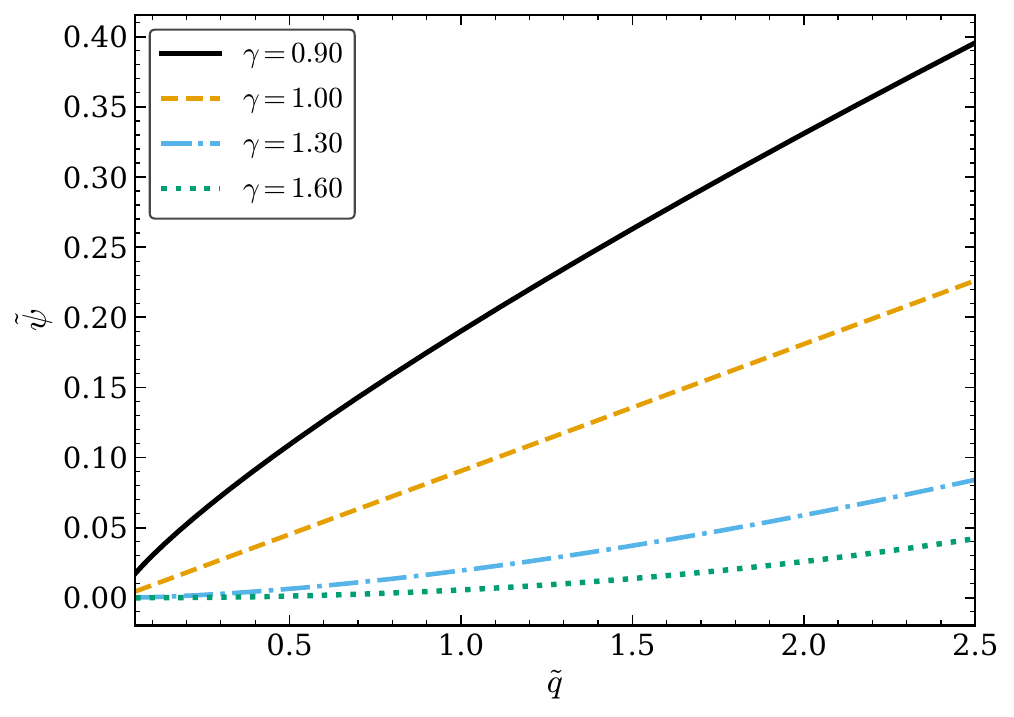}
  \caption{Yang--Mills potential $\tilde{\psi}$ versus $\tilde{q}$ for $\gamma=0.90,1.00,1.30,1.60$ at $S=6$, $C=4$, $\tilde{Q}=1$, $R=2$, from Eq.~\eqref{eq:psi}.}
  \label{fig:psi_q}
\end{figure}

\begin{figure}[htbp]
  \centering
  \includegraphics[width=0.52\textwidth]{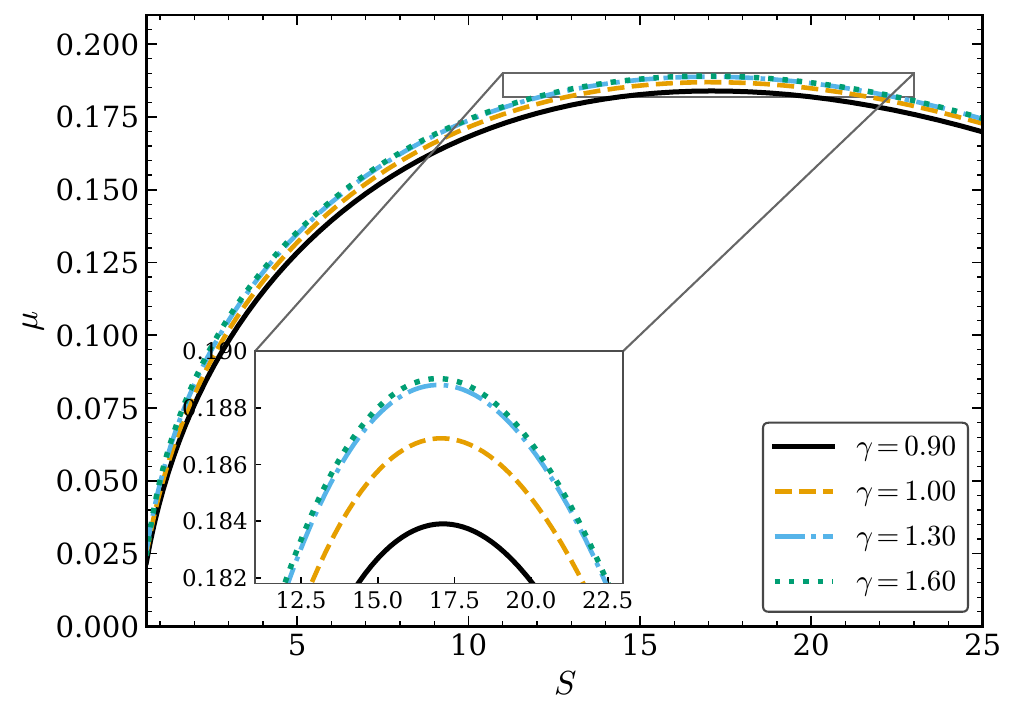}
  \caption{Chemical potential $\mu$ conjugate to the central charge versus entropy $S$ for $\gamma=0.90,1.00,1.30,1.60$ at $C=4$, $\tilde{Q}=1$, $\tilde{q}=0.8$, $R=2$, from Eq.~\eqref{eq:mu}. The inset magnifies the peak region $S\in[11,23]$, where the curves order by $\gamma$.}
  \label{fig:mu_S}
\end{figure}

The temperature, the Yang--Mills potential, and the chemical potential are shown in Figs.~\ref{fig:T_S}, \ref{fig:psi_q}, and~\ref{fig:mu_S} as functions of their natural arguments. Start with the temperature. In Fig.~\ref{fig:T_S} it climbs out of a small-entropy regime, where the Yang--Mills and charge terms of Eq.~\eqref{eq:T} dominate, and settles onto the large-entropy branch fixed by the $\sqrt{C}$ and $S^{1/2}$ terms. Past $S\simeq3$ the four curves crowd into a single line width, which is why the inset magnifies the minimum near $S\in[8,20]$. The ordering there is not arbitrary: the leading Yang--Mills piece scales as $S^{1/2-2\gamma}$, so a larger $\gamma$ depresses $\tilde{T}$ at fixed entropy.

At a $\gamma$-dependent entropy the temperature of Fig.~\ref{fig:T_S} crosses zero, and that crossing is the extremal CFT state. Its location shifts with $\gamma$ through the first term of Eq.~\eqref{eq:T}, whose weight $2^{-\gamma-2}\pi^{2\gamma-3/2}$ and exponent $\tfrac{1}{2}-2\gamma$ both carry the nonlinearity. Setting $\gamma=1$ collapses the temperature to the power-Maxwell form, a useful check against the literature on charged AdS thermodynamics~\cite{75,81}.

In Fig.~\ref{fig:psi_q} the Yang--Mills potential grows with $\tilde{q}$ and steepens as $\gamma$ increases. The factor $\tilde{q}^{2\gamma-1}$ in Eq.~\eqref{eq:psi} drives this, since a larger exponent raises both the power of the charge and the prefactor $\gamma/(4\gamma-3)$. As $\tilde{q}\to0$ the potential vanishes for every $\gamma>1/2$, which returns the pure Maxwell--AdS case with no Yang--Mills conjugate.

The chemical potential of Fig.~\ref{fig:mu_S} starts positive at small entropy, turns over, and falls as $S$ grows. Two terms in Eq.~\eqref{eq:mu} fight over its sign, the rising $\sqrt{S}$ term and the falling $S^{3/2}/C^{3/2}$ term, and their balance sets $\mu$ across the plotted range. Raising $\gamma$ pushes the curve down, acting through the Yang--Mills contribution, the only $\gamma$-dependent term in Eq.~\eqref{eq:mu}. The sign matters. Since $\mu$ is conjugate to the central charge in the first law, it decides whether enlarging the dual CFT costs or releases energy, and that enters the censorship analysis once $dC\ne0$.

\begin{table*}[htbp]
  \centering
  \setlength{\tabcolsep}{6pt}
  \renewcommand{\arraystretch}{1.25}
  \begin{tabularx}{\textwidth}{C C C C C C C C}
    \toprule
    \rowcolor{orange!50}
    $S$ & $C$ & $\tilde{Q}$ & $\tilde{q}$ & $\gamma$ & $\tilde{T}$ & $\tilde{\psi}$ & $\mu$ \\
    \midrule
    4  & 4 & 1.0 & 0.8 & 1.0 & 0.163365 & 0.088623 & 0.118468 \\
    6  & 4 & 1.2 & 0.9 & 1.3 & 0.149936 & 0.016413 & 0.142850 \\
    9  & 6 & 1.5 & 1.0 & 1.6 & 0.122024 & 0.001446 & 0.117096 \\
    12 & 8 & 1.8 & 1.2 & 0.9 & 0.100730 & 0.090270 & 0.095356 \\
    \bottomrule
  \end{tabularx}
  \caption{Boundary equations of state evaluated from Eqs.~\eqref{eq:T}, \eqref{eq:psi}, and~\eqref{eq:mu} at four representative points (geometric units, $R$ taken as $2,2,2.5,3$ respectively). The electric potential $\tilde{\phi}$ from Eq.~\eqref{eq:phi} is omitted for brevity and is cross-checked in Table~\ref{tab:eos_verify}.}
  \label{tab:eos_values}
\end{table*}

Four working points, used throughout the paper, are collected in Table~\ref{tab:eos_values}. The temperature stays positive in every row, since each point sits on the large-entropy branch of Fig.~\ref{fig:T_S}, well away from the extremal crossing. Between the $\gamma=1.3$ and $\gamma=1.6$ rows the Yang--Mills potential drops sharply; this traces to the $S^{3/2-2\gamma}$ scaling of Eq.~\eqref{eq:psi}, which suppresses $\tilde{\psi}$ at large entropy and large exponent.

\subsection{Euler relation and the conformal equation of state}\label{isec2d}

The internal energy~\eqref{eq:E} is a sum of terms, each homogeneous in $(S,C,\tilde{Q},\tilde{q})$ at fixed $R$. Adding the weighted conjugates therefore reconstructs $E$ exactly. Using the closed forms~\eqref{eq:T}--\eqref{eq:mu} together with the definition~\eqref{eq:mu_def}, one finds the Euler relation
\begin{equation}
E=\tilde{T}S+\tilde{\phi}\tilde{Q}+\frac{3\gamma-1}{2\gamma}\,\tilde{\psi}\tilde{q}+\mu C.
\label{eq:euler}
\end{equation}
The Yang--Mills term carries the weight $(3\gamma-1)/(2\gamma)$ rather than unity, which reflects the scaling dimension that the exponent $\gamma$ assigns to the non-Abelian charge in Eq.~\eqref{eq:E}. Relation~\eqref{eq:euler} is the integrated form of the first law~\eqref{eq:firstlaw} for this homogeneous energy, and it is the boundary analogue of a Smarr formula. Together with the conformal equation of state,
\begin{equation}
p\,\Vol=\frac{E}{2},
\label{eq:conformal}
\end{equation}
it fixes the thermodynamics with no free integration constant.

We checked Eqs.~\eqref{eq:T}--\eqref{eq:mu}, the Euler relation~\eqref{eq:euler}, and the conformal equation of state~\eqref{eq:conformal} by recomputing every conjugate as a numerical derivative of $E$ and comparing with the closed forms. Table~\ref{tab:eos_verify} reports the temperature, electric potential, Yang--Mills potential, and chemical potential against the corresponding partial derivatives of $E$, while Table~\ref{tab:euler_verify} reports the two integrated relations. The agreement holds to the precision of the finite-difference scheme, which confirms that the equations of state are mutually consistent and that the Euler relation~\eqref{eq:euler} is an exact identity of the model rather than an approximation.

\begin{table*}[htbp]
  \centering
  \setlength{\tabcolsep}{5pt}
  \renewcommand{\arraystretch}{1.25}
  \begin{tabularx}{\textwidth}{C C C C C C C}
    \toprule
    \rowcolor{orange!50}
    $\gamma$ & $\tilde{T}$ (Eq.~\ref{eq:T}) & $\partial E/\partial S$ & $\tilde{\phi}$ (Eq.~\ref{eq:phi}) & $\partial E/\partial \tilde{Q}$ & $\tilde{\psi}$ (Eq.~\ref{eq:psi}) & $\partial E/\partial \tilde{q}$ \\
    \midrule
    1.0 & 0.163365 & 0.163365 & 0.110778 & 0.110778 & 0.088623 & 0.088623 \\
    1.3 & 0.149936 & 0.149936 & 0.108540 & 0.108540 & 0.016413 & 0.016413 \\
    1.6 & 0.122024 & 0.122024 & 0.072360 & 0.072360 & 0.001446 & 0.001446 \\
    0.9 & 0.100730 & 0.100730 & 0.054270 & 0.054270 & 0.090270 & 0.090270 \\
    \bottomrule
  \end{tabularx}
  \caption{Closed-form equations of state versus numerical derivatives of $E$ at the four points of Table~\ref{tab:eos_values}. Each conjugate matches the corresponding partial derivative of Eq.~\eqref{eq:E} to the resolution of the difference scheme (relative deviation below $10^{-9}$).}
  \label{tab:eos_verify}
\end{table*}

\begin{table*}[htbp]
  \centering
  \setlength{\tabcolsep}{6pt}
  \renewcommand{\arraystretch}{1.25}
  \begin{tabularx}{\textwidth}{C C C C C C}
    \toprule
    \rowcolor{orange!50}
    $S$ & $\gamma$ & $E$ (Eq.~\ref{eq:E}) & RHS of Eq.~\eqref{eq:euler} & $p\,\Vol$ & $E/2$ \\
    \midrule
    4  & 1.0 & 1.30901 & 1.30901 & 0.654505 & 0.654505 \\
    6  & 1.3 & 1.61774 & 1.61774 & 0.808872 & 0.808872 \\
    9  & 1.6 & 1.91105 & 1.91105 & 0.955524 & 0.955524 \\
    12 & 0.9 & 2.17161 & 2.17161 & 1.085800 & 1.085800 \\
    \bottomrule
  \end{tabularx}
  \caption{Verification of the Euler relation~\eqref{eq:euler} and the conformal equation of state~\eqref{eq:conformal} at the four points of Table~\ref{tab:eos_values}. The reconstructed energy $\tilde{T}S+\tilde{\phi}\tilde{Q}+\tfrac{3\gamma-1}{2\gamma}\tilde{\psi}\tilde{q}+\mu C$ reproduces $E$, and $p\,\Vol$ reproduces $E/2$, in both cases to all digits shown.}
  \label{tab:euler_verify}
\end{table*}

The match in Table~\ref{tab:euler_verify} is exact to the digits shown because Eqs.~\eqref{eq:euler} and~\eqref{eq:conformal} are algebraic identities of the homogeneous energy~\eqref{eq:E}, not numerical coincidences. The first follows from Euler's theorem applied term by term, since the four contributions to $E$ carry definite weights in $S$, $C$, $\tilde{Q}^{2}$, and $\tilde{q}^{2\gamma}$ that add to unity once the Yang--Mills factor $(3\gamma-1)/(2\gamma)$ is included. The second is the statement that a CFT on a sphere has traceless stress tensor, so the energy and pressure are locked by $p=E/(2\Vol)$. Read together with Table~\ref{tab:eos_verify}, these checks establish that the thermodynamic description used in the rest of the paper is internally closed.

\subsection{Local thermodynamic stability}\label{isec2e}

\begin{figure}[htbp]
  \centering
  \includegraphics[width=0.52\textwidth]{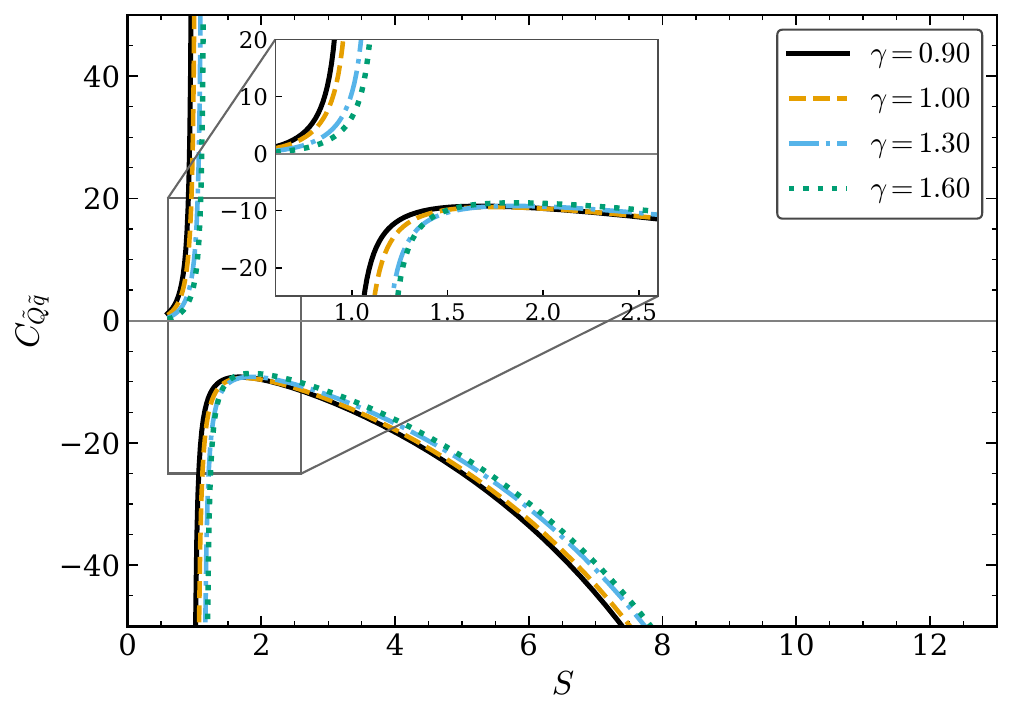}
  \caption{Heat capacity $C_{\tilde{Q}\tilde{q}}$ from Eq.~\eqref{eq:heatcap} versus entropy $S$ for $\gamma=0.90,1.00,1.30,1.60$ at $C=4$, $\tilde{Q}=1$, $\tilde{q}=0.8$, $R=2$. The divergence near $S\simeq1$ is a Davies point separating the small-$S$ stable branch from the large-$S$ unstable branch. The inset magnifies the sign-change region $S\in[0.6,2.6]$.}
  \label{fig:heatcap}
\end{figure}

Local stability is read from the heat capacity at fixed charges and central charge,
\begin{equation}
C_{\tilde{Q}\tilde{q}}=\tilde{T}\left(\frac{\partial S}{\partial \tilde{T}}\right)_{\tilde{Q},\tilde{q},C,\Vol}=\frac{\tilde{T}}{(\partial \tilde{T}/\partial S)}.
\label{eq:heatcap}
\end{equation}
A positive $C_{\tilde{Q}\tilde{q}}$ marks a locally stable branch, a negative value an unstable one, and a divergence a Davies point where $\partial\tilde{T}/\partial S=0$. Figure~\ref{fig:heatcap} plots $C_{\tilde{Q}\tilde{q}}$ against $S$ for four exponents.

In Fig.~\ref{fig:heatcap} the heat capacity diverges near $S\simeq1$ and flips sign. That divergence is the Davies condition itself: $\partial\tilde{T}/\partial S$ passes through zero at the turning point of the temperature in Fig.~\ref{fig:T_S}. Left of the divergence the capacity is positive, so the configuration is locally stable; right of it the capacity is negative, the mark of the Schwarzschild-like branch on which adding energy lowers the temperature. As $\gamma$ grows the Davies point drifts upward in $S$, from $S\simeq0.979$ at $\gamma=0.9$ to $S\simeq1.162$ at $\gamma=1.6$, tracking the $\gamma$-dependence of the leading term of Eq.~\eqref{eq:T} and matching Table~\ref{tab:heatcap}.

\begin{table*}[htbp]
  \centering
  \setlength{\tabcolsep}{7pt}
  \renewcommand{\arraystretch}{1.25}
  \begin{tabularx}{\textwidth}{C C C C}
    \toprule
    \rowcolor{orange!50}
    $\gamma$ & $S_{\mathrm{Davies}}$ & $C_{\tilde{Q}\tilde{q}}(S=2)$ & $C_{\tilde{Q}\tilde{q}}(S=10)$ \\
    \midrule
    0.90 & 0.97946 & $-9.6179$ & $-106.91$ \\
    1.00 & 1.02927 & $-9.6411$ & $-102.07$ \\
    1.30 & 1.12294 & $-9.2714$ & $-94.859$ \\
    1.60 & 1.16159 & $-8.7125$ & $-92.586$ \\
    \bottomrule
  \end{tabularx}
  \caption{Davies point $S_{\mathrm{Davies}}$ where $C_{\tilde{Q}\tilde{q}}$ diverges, and the heat capacity at two reference entropies, for $C=4$, $\tilde{Q}=1$, $\tilde{q}=0.8$, $R=2$. The Davies entropy increases with $\gamma$, while the large-entropy branch stays negative.}
  \label{tab:heatcap}
\end{table*}

The drift of the Davies point and the persistence of the unstable branch are quantified in Table~\ref{tab:heatcap}. At both sampled entropies beyond the divergence the heat capacity is negative, so the large black hole in this ensemble is locally unstable, exactly as Schwarzschild--AdS is in the same fixed-charge frame. The Davies entropy rises monotonically with $\gamma$ because the Yang--Mills term of Eq.~\eqref{eq:T} shifts the temperature turning point, and for the same reason the negative capacity weakens slightly as $\gamma$ grows. This local picture complements the global censorship test of Sec.~\ref{isec4}. Stability of the ensemble and survival of the horizon under perturbation are distinct questions, and the model behaves well on both within the ranges studied.

\subsection{On-shell free energy}\label{isec2f}

The phase behavior is read most directly from the on-shell free energy. In the fixed-charge ensemble the relevant potential is
\begin{equation}
\mathcal{F}=E-\tilde{T}S,
\label{eq:freeenergy}
\end{equation}
which is the boundary analogue of the Gibbs energy at fixed $\tilde{Q}$, $\tilde{q}$, $C$, and $\Vol$. Using the Euler relation~\eqref{eq:euler} to eliminate $E$, the free energy can be written as
\begin{equation}
\mathcal{F}=\tilde{\phi}\tilde{Q}+\frac{3\gamma-1}{2\gamma}\tilde{\psi}\tilde{q}+\mu C,
\label{eq:F_euler}
\end{equation}
so the entropic term $\tilde{T}S$ has been traded for the charge and central-charge contributions. The sign of $\mathcal{F}$ decides whether the black-hole phase is favored over the radiation phase at the same temperature: a negative $\mathcal{F}$ signals the dominant phase, and the locus $\mathcal{F}=0$ is the boundary analogue of a Hawking--Page transition. Because $\mu$ turns over and changes sign in Fig.~\ref{fig:mu_S}, while $\tilde{\phi}\tilde{Q}$ and $\tilde{\psi}\tilde{q}$ stay positive, the central-charge term is what can push $\mathcal{F}$ through zero at large entropy. The Yang--Mills contribution enters with the same weight $(3\gamma-1)/(2\gamma)$ it carries in Eq.~\eqref{eq:euler}, so a larger exponent raises the free-energy cost of the non-Abelian charge and shifts the transition. This free-energy structure sits behind the heat-capacity sign change of Fig.~\ref{fig:heatcap}: the Davies divergence marks the entropy where $\partial\tilde{T}/\partial S$ vanishes, which is also where the slope of $\mathcal{F}(\tilde{T})$ becomes singular and the locally stable branch ends.

\section{Charged massive scalar field and horizon fluxes}\label{isec3}

To probe the \EMPYM\ black hole we introduce a charged massive scalar field $\Psi$ with mass $\mu_{s}$ and electric charge $q_{s}$, not to be confused with the Yang--Mills parameter $q$, minimally coupled to gravity and to the Maxwell field. Its equation of motion is
\begin{equation}
(\nabla_{\mu}-iq_{s}A_{\mu})(\nabla^{\mu}-iq_{s}A^{\mu})\Psi-\mu_{s}^{2}\Psi=0,
\label{eq:KG}
\end{equation}
where $A_{\mu}$ is the Maxwell gauge potential. In explicit form,
\begin{equation}
\frac{1}{\sqrt{-g}}(\partial_{\mu}-iq_{s}A_{\mu})\big[\sqrt{-g}\,g^{\mu\nu}(\partial_{\nu}-iq_{s}A_{\nu})\Psi\big]-\mu_{s}^{2}\Psi=0.
\label{eq:KG2}
\end{equation}
Because the spacetime is static and spherically symmetric, we separate the field as
\begin{equation}
\Psi(t,r,\theta,\phi)=e^{-i\omega t}R_{lm}(r)Y_{lm}(\theta,\phi),
\label{eq:sep}
\end{equation}
with $\omega$ the frequency, $R_{lm}(r)$ the radial function, and $Y_{lm}(\theta,\phi)$ the spherical harmonics. The separation constant $l(l+1)$ comes from the angular equation. The angular part is standard and normalized, so we focus on the radial component. Substituting Eq.~\eqref{eq:sep} into Eq.~\eqref{eq:KG} gives the radial equation
\begin{equation}
\frac{1}{r^{2}}\frac{d}{dr}\!\left[r^{2}f(r)\frac{dR_{lm}}{dr}\right]+\left[\frac{(\omega-q_{s}\phi(r))^{2}}{f(r)}-\frac{l(l+1)}{r^{2}}-\mu_{s}^{2}\right]R_{lm}=0,
\label{eq:radial}
\end{equation}
where $\phi(r)=Q/r$ is the electric potential and $f(r)$ is the \EMPYM\ metric function, which carries the Maxwell charge $Q$, the Yang--Mills parameter $q$, and the exponent $\gamma$ through the term $\Qcal\,r^{-(4\gamma-2)}$. Introducing the tortoise coordinate $r_{*}$ through $dr/dr_{*}=f(r)$, the radial equation becomes
\begin{equation}
\frac{d^{2}R_{lm}}{dr_{*}^{2}}+\frac{2f(r)}{r}\frac{dR_{lm}}{dr_{*}}+\left[\left(\omega-\frac{q_{s}Q}{r}\right)^{2}-f(r)\left(\frac{l(l+1)}{r^{2}}+\mu_{s}^{2}\right)\right]R_{lm}=0.
\label{eq:radial_tort}
\end{equation}
As $r$ runs from the horizon $r_{h}$ to infinity, $r_{*}$ covers $(-\infty,+\infty)$. Near the horizon $f(r)\to0$ and $r\to r_{h}$, so the radial equation reduces to
\begin{equation}
\frac{d^{2}R_{lm}}{dr_{*}^{2}}+\left(\omega-\frac{q_{s}Q}{r_{h}}\right)^{2}R_{lm}=0,
\label{eq:nearhor}
\end{equation}
and with the horizon potential $\phi_{h}=Q/r_{h}$ this is
\begin{equation}
\frac{d^{2}R_{lm}}{dr_{*}^{2}}+(\omega-q_{s}\phi_{h})^{2}R_{lm}=0.
\label{eq:nearhor2}
\end{equation}
The general solution is a pair of plane waves, $R_{lm}\sim\exp[\pm i(\omega-q_{s}\phi_{h})r_{*}]$. The negative sign is the incoming wave absorbed by the black hole, so near the horizon
\begin{equation}
\Psi\sim e^{-i(\omega-q_{s}\phi_{h})r_{*}}e^{-i\omega t}Y_{lm}(\theta,\phi).
\label{eq:nearhor_sol}
\end{equation}

The energy and charge carried into the hole follow from the stress tensor and the current. For the charged scalar,
\begin{equation}
T^{\mu}{}_{\nu}=\frac{1}{2}D^{\mu}\Psi\,\partial_{\nu}\Psi^{*}+\frac{1}{2}D^{*\mu}\Psi^{*}\,\partial_{\nu}\Psi-\delta^{\mu}_{\nu}\left[\frac{1}{2}D_{\alpha}\Psi\,D^{*\alpha}\Psi^{*}-\frac{1}{2}\mu_{s}^{2}\Psi\Psi^{*}\right],
\label{eq:Tmunu}
\end{equation}
with $D_{\mu}=\partial_{\mu}-iq_{s}A_{\mu}$, and the current is
\begin{equation}
j^{\mu}=-\frac{i}{2}q_{s}(\Psi^{*}D^{\mu}\Psi-\Psi D^{*\mu}\Psi^{*}).
\label{eq:current}
\end{equation}
Integrating $T^{r}{}_{t}$ over the horizon gives the energy flux, and the radial current gives the charge flux:
\begin{equation}
\frac{dE}{dt}=\int_{\mathcal{H}}T^{r}{}_{t}\sqrt{-g}\,d\theta\,d\phi,\qquad \frac{dQ}{dt}=-\int_{\mathcal{H}}j^{r}\sqrt{-g}\,d\theta\,d\phi.
\label{eq:flux_int}
\end{equation}
Substituting the near-horizon solution~\eqref{eq:nearhor_sol} and using the unit normalization of the spherical harmonics, the angular integral contributes the horizon area factor $r_{h}^{2}$, and the radial derivative of the plane wave brings down $(\omega-q_{s}\phi_{h})$. The two fluxes reduce to
\begin{equation}
\frac{dE}{dt}=\omega(\omega-q_{s}\phi_{h})r_{h}^{2},\qquad \frac{dQ}{dt}=q_{s}(\omega-q_{s}\phi_{h})r_{h}^{2}.
\label{eq:fluxes}
\end{equation}
Their ratio $q_{s}/\omega$ is independent of the geometry, a feature that traces back to the structure of Eq.~\eqref{eq:nearhor_sol}, where the energy and charge are carried by the same phase $(\omega-q_{s}\phi_{h})r_{*}$. When $\omega>q_{s}\phi_{h}$ both fluxes are positive and the field deposits energy and charge into the hole; when $\omega<q_{s}\phi_{h}$ both are negative and the field extracts them, which is superradiance. Over an interval $dt$ the black hole parameters change by
\begin{equation}
dM=\omega(\omega-\tilde{q}_{s}\tilde{\phi}_{h})r_{h}^{2}\,dt,\qquad d\tilde{Q}=\tilde{q}_{s}(\omega-\tilde{q}_{s}\tilde{\phi}_{h})r_{h}^{2}\,dt+\frac{\tilde{Q}}{2C}\,dC,
\label{eq:dMdQ}
\end{equation}
where the second term in $d\tilde{Q}$ accounts for the variation of the central charge independently of the scalar absorption, via $\tilde{Q}=2Q\sqrt{C}$. The Yang--Mills charge $\tilde{q}$ and the exponent $\gamma$ enter through the extended first law, with their variations carried by $\tilde{\psi}$ and $\mu$.

The superradiant threshold $\omega_{c}=\tilde{q}_{s}\tilde{\phi}_{h}$ separates absorption from extraction, and Table~\ref{tab:superrad} lists it for several horizon radii and scalar charges. It grows linearly with the scalar charge $\tilde{q}_{s}$, since the boundary potential $\tilde{\phi}_{h}=\tilde{Q}/r_{h}$ multiplies it directly in Eq.~\eqref{eq:fluxes}, and it falls as the horizon expands, because a larger $r_{h}$ dilutes that potential. The near-extremal entry with $r_{h}=1.4977$, taken from the single-horizon row of Table~\ref{tab:horizon}, sits between the small and large reference radii.

\begin{table*}[htbp]
  \centering
  \setlength{\tabcolsep}{7pt}
  \renewcommand{\arraystretch}{1.25}
  \begin{tabularx}{\textwidth}{C C C C C}
    \toprule
    \rowcolor{orange!50}
    $r_{h}$ & $\tilde{Q}$ & $\tilde{\phi}_{h}=\tilde{Q}/r_{h}$ & $\tilde{q}_{s}$ & $\omega_{c}=\tilde{q}_{s}\tilde{\phi}_{h}$ \\
    \midrule
    1.0000 & 1.0 & 1.00000 & 0.6 & 0.60000 \\
    1.0000 & 1.0 & 1.00000 & 1.0 & 1.00000 \\
    1.0000 & 1.0 & 1.00000 & 1.5 & 1.50000 \\
    1.4000 & 1.0 & 0.71429 & 0.6 & 0.42857 \\
    1.4000 & 1.0 & 0.71429 & 1.0 & 0.71429 \\
    2.0000 & 1.2 & 0.60000 & 1.5 & 0.90000 \\
    1.4977 & 0.9 & 0.60092 & 0.6 & 0.36055 \\
    1.4977 & 0.9 & 0.60092 & 1.5 & 0.90138 \\
    \bottomrule
  \end{tabularx}
  \caption{Superradiant threshold $\omega_{c}=\tilde{q}_{s}\tilde{\phi}_{h}$ for a charged scalar with boundary charge $\tilde{q}_{s}$ scattering off horizons of radius $r_{h}$ and boundary charge $\tilde{Q}$ (geometric units). Modes with $\omega<\omega_{c}$ extract energy and charge; modes with $\omega>\omega_{c}$ are absorbed.}
  \label{tab:superrad}
\end{table*}

\section{Weak cosmic censorship under scalar absorption}\label{isec4}

We now test the \WCCC\ by following the metric function after absorption. The horizon is the largest real root of $f(r)=0$. For an extremal black hole $f$ has a double root, $f_{\min}=0$ and $r_{\min}=r_{h}$. For a near-extremal black hole the minimum is slightly negative, $f_{\min}=\delta<0$ with $|\delta|\ll1$, and $r_{h}=r_{\min}+\varepsilon$ with $0<\varepsilon\ll1$. After absorption the parameters move from $(M,\tilde{Q},\gamma,C)$ to $(M+dM,\tilde{Q}+d\tilde{Q},\gamma+d\gamma,C+dC)$, and the minimum shifts to $r_{\min}+dr_{\min}$. The conjecture survives if $f_{\min}\le0$ in the final state and fails if $f_{\min}>0$, which would expose a naked singularity. The new minimum is
\begin{equation}
f_{\mathrm{final}}=f(r_{\min}+dr_{\min})+\frac{\partial f}{\partial M}dM+\frac{\partial f}{\partial \tilde{Q}}d\tilde{Q}+\frac{\partial f}{\partial \gamma}d\gamma+\frac{\partial f}{\partial C}dC.
\label{eq:ffinal0}
\end{equation}
Because $f'(r_{\min})=0$, the term proportional to $dr_{\min}$ does not contribute at first order, so
\begin{equation}
f_{\mathrm{final}}=f_{\min}+\left.\frac{\partial f}{\partial M}\right|_{r_{\min}}dM+\left.\frac{\partial f}{\partial \tilde{Q}}\right|_{r_{\min}}d\tilde{Q}+\left.\frac{\partial f}{\partial \gamma}\right|_{r_{\min}}d\gamma+\left.\frac{\partial f}{\partial C}\right|_{r_{\min}}dC.
\label{eq:ffinal}
\end{equation}
The partial derivatives at the minimum are
\begin{equation}
\left.\frac{\partial f}{\partial M}\right|_{r_{\min}}=-\frac{2}{r_{\min}},\quad
\left.\frac{\partial f}{\partial \tilde{Q}}\right|_{r_{\min}}=\frac{2\tilde{Q}}{r_{\min}^{2}},\quad
\left.\frac{\partial f}{\partial \gamma}\right|_{r_{\min}}=-\frac{\partial \Qcal}{\partial \gamma}\frac{1}{r_{\min}^{4\gamma-2}}+\frac{4\Qcal\ln r_{\min}}{r_{\min}^{4\gamma-2}},\quad
\left.\frac{\partial f}{\partial C}\right|_{r_{\min}}=-\frac{r_{\min}^{2}}{\ell^{2}C},
\label{eq:partials}
\end{equation}
where we used $\ell=2\sqrt{C}$ from $C=\ell^{2}/4$. The exponent derivative of the Yang--Mills amplitude is
\begin{equation}
\frac{\partial \Qcal}{\partial \gamma}=\frac{(2q^{2})^{\gamma}\ln(2q^{2})}{2(4\gamma-3)}-\frac{2(2q^{2})^{\gamma}}{(4\gamma-3)^{2}}.
\label{eq:dQdg}
\end{equation}
The mass and charge variations come from the fluxes,
\begin{equation}
dM=\omega(\omega-\tilde{q}_{s}\tilde{\phi}_{h})r_{h}^{2}\,dt,\qquad d\tilde{Q}=\tilde{q}_{s}(\omega-\tilde{q}_{s}\phi_{h})r_{h}^{2}\,dt,
\label{eq:dMdQ2}
\end{equation}
with $\phi_{h}=\tilde{Q}/r_{h}$. The variations $d\gamma$ and $dC$ are independent and are treated as parameters.

\subsection{Horizon shift and the second law}\label{isec4shift}

Before specializing to the extremal and near-extremal cases it is worth recording how the horizon itself responds, since this controls the entropy balance. Requiring $f=0$ at the horizon before and after absorption, and varying all parameters,
\begin{equation}
df=\frac{\partial f}{\partial M}dM+\frac{\partial f}{\partial \tilde{Q}}d\tilde{Q}+\frac{\partial f}{\partial r_{h}}dr_{h}+\frac{\partial f}{\partial \gamma}d\gamma+\frac{\partial f}{\partial C}dC=0,
\label{eq:df_horizon}
\end{equation}
where the radial derivative is the slope $f'(r_{h})$ given in Eq.~\eqref{eq:fprime}. Solving for the horizon displacement,
\begin{equation}
dr_{h}=-\frac{1}{f'(r_{h})}\left(\frac{\partial f}{\partial M}dM+\frac{\partial f}{\partial \tilde{Q}}d\tilde{Q}+\frac{\partial f}{\partial \gamma}d\gamma+\frac{\partial f}{\partial C}dC\right).
\label{eq:drh}
\end{equation}
The entropy is $\tilde{S}=S=\pi r_{h}^{2}$ from Eq.~\eqref{eq:Sentropy}, so its variation is
\begin{equation}
d\tilde{S}=2\pi r_{h}\,dr_{h}.
\label{eq:dS}
\end{equation}
Away from extremality the slope $f'(r_{h})$ is strictly positive at the outer horizon, so the sign of $dr_{h}$ follows from the bracket in Eq.~\eqref{eq:drh}. For an absorbed mode with $\omega>\tilde{q}_{s}\tilde{\phi}_{h}$ both $dM$ and $d\tilde{Q}$ are positive, and the dominant $-\tfrac{2}{r_{h}}dM$ contribution drives $dr_{h}>0$, so the horizon grows and $d\tilde{S}\ge0$. This is the second law for the absorption process, and it holds in the \EMPYM\ background for the same reason it holds in the Reissner--Nordstr\"om case: the energy delivered through the horizon exceeds the work done against the electric potential. The Yang--Mills and central-charge terms in Eq.~\eqref{eq:drh} enter only through $d\gamma$ and $dC$, so when those are held fixed the entropy balance is controlled entirely by the scalar flux.

\subsection{Extremal case}\label{isec4a}

For an extremal black hole $r_{\min}=r_{h}$ and $f_{\min}=0$. Substituting into Eq.~\eqref{eq:ffinal},
\begin{equation}
f_{\mathrm{final}}=-\frac{2}{r_{h}}dM+\frac{2\tilde{Q}}{r_{h}^{2}}d\tilde{Q}+\left.\frac{\partial f}{\partial \gamma}\right|_{r_{h}}d\gamma+\left.\frac{\partial f}{\partial C}\right|_{r_{h}}dC.
\label{eq:ffinal_ext}
\end{equation}
Inserting $dM$ and $d\tilde{Q}$ and using $\phi_{h}=\tilde{Q}/r_{h}$,
\begin{equation}
f_{\mathrm{final}}=2(\omega-\tilde{q}_{s}\phi_{h})(\tilde{q}_{s}\tilde{\phi}_{h}-\omega)r_{h}\,dt+\left.\frac{\partial f}{\partial \gamma}\right|_{r_{h}}d\gamma+\left.\frac{\partial f}{\partial C}\right|_{r_{h}}dC,
\label{eq:ffinal_ext2}
\end{equation}
and the first term reduces to
\begin{equation}
f_{\mathrm{final}}\big|_{d\gamma=dC=0}=-2(\omega-\tilde{q}_{s}\tilde{\phi}_{h})^{2}r_{h}\,dt\le0.
\label{eq:ffinal_ext3}
\end{equation}
The scalar contribution alone is therefore non-positive, vanishing only at the superradiant threshold $\omega=\tilde{q}_{s}\tilde{\phi}_{h}$. The extremal hole stays extremal or becomes slightly sub-extremal with a negative minimum, so the horizon survives. Figure~\ref{fig:wccc} shows $f_{\mathrm{final}}/dt$ for three scalar charges.

Across the whole frequency range the curves of Fig.~\ref{fig:wccc} stay at or below zero, which is just the graphical content of Eq.~\eqref{eq:ffinal_ext3}. Each parabola touches zero at its own threshold $\omega=\tilde{q}_{s}\tilde{\phi}_{h}$ and nowhere else, because there the energy and charge fluxes balance and the leading shift of the minimum cancels. The reason is structural: combining the flux expressions~\eqref{eq:dMdQ2} in Eq.~\eqref{eq:ffinal_ext2} leaves a perfect square $(\omega-\tilde{q}_{s}\tilde{\phi}_{h})^{2}$, and no real frequency can turn that positive. For the \EMPYM\ family this reproduces the protection of extremal black holes against test fields established by Nat\'ario, Queimada, and Vicente~\cite{Natario:2016bay}, now with an explicit Yang--Mills sector.

If $d\gamma$ and $dC$ are allowed, the sign of $f_{\mathrm{final}}$ depends on their size. For variations of order $dt$ or smaller the negative scalar term dominates and the conjecture holds. Only large, externally imposed changes of $\gamma$ or $C$ could overturn the inequality, and these are not produced by scalar absorption, since $\gamma$ is a fixed coupling of the Yang--Mills Lagrangian and $C$ is a fixed property of the dual CFT.

\subsection{Near-extremal case}\label{isec4b}

Let the initial hole be near-extremal,
\begin{equation}
r_{h}=r_{\min}+\varepsilon,\qquad f_{\min}=\delta<0,\qquad |\delta|\ll1,\qquad 0<\varepsilon\ll1.
\label{eq:nearext}
\end{equation}
Expanding $f$ about $r_{\min}$ and using $f(r_{h})=0$,
\begin{equation}
\delta=-\frac{1}{2}f''(r_{\min})\varepsilon^{2}+O(\varepsilon^{3}),
\label{eq:delta}
\end{equation}
so $\delta$ is negative and of order $\varepsilon^{2}$. The horizon potential expands as
\begin{equation}
\tilde{\phi}_{h}=\frac{\tilde{Q}}{r_{\min}+\varepsilon}=\tilde{\phi}_{\min}\left(1-\frac{\varepsilon}{r_{\min}}+O(\varepsilon^{2})\right),
\label{eq:phih_exp}
\end{equation}
with $\tilde{\phi}_{\min}=\tilde{Q}/r_{\min}$, and $r_{h}^{2}=r_{\min}^{2}+2r_{\min}\varepsilon+O(\varepsilon^{2})$. Carrying the expansion through Eq.~\eqref{eq:ffinal},
\begin{equation}
f_{\mathrm{final}}=\delta-2(\omega-\tilde{q}_{s}\tilde{\phi}_{\min})^{2}r_{\min}\,dt+\left.\frac{\partial f}{\partial \gamma}\right|_{r_{\min}}d\gamma+\left.\frac{\partial f}{\partial C}\right|_{r_{\min}}dC+O(\varepsilon^{2}dt,\varepsilon\,d\gamma,\varepsilon\,dC,dt^{2}).
\label{eq:ffinal_near}
\end{equation}
If $dt$ is of order $\varepsilon$, the term $-2(\omega-\tilde{q}_{s}\tilde{\phi}_{\min})^{2}r_{\min}\,dt$ is of order $\varepsilon$, while $\delta$ is of order $\varepsilon^{2}$. For generic frequencies $\omega\ne\tilde{q}_{s}\tilde{\phi}_{\min}$ the linear term then sets the sign, and since it is non-positive, $f_{\mathrm{final}}<0$ and the horizon persists. At the tuned frequency $\omega=\tilde{q}_{s}\tilde{\phi}_{\min}$ the linear term vanishes and the order-$\varepsilon^{2}$ contributions decide; with $d\gamma$ and $dC$ small, the negative $\delta$ keeps $f_{\mathrm{final}}<0$.

\begin{figure}[htbp]
  \centering
  \begin{subfigure}[t]{0.49\textwidth}
    \centering
    \includegraphics[width=\textwidth]{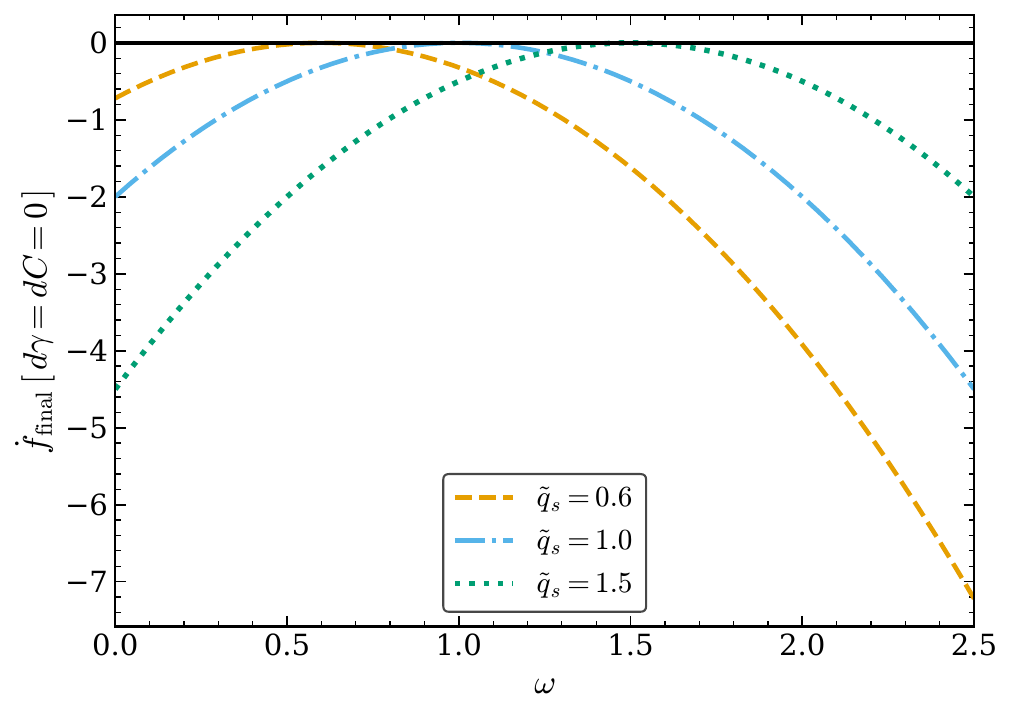}
    \caption{Leading change of the extremal minimum, $\dot{f}_{\mathrm{final}}=-2(\omega-\tilde{q}_{s}\tilde{\phi}_{h})^{2}r_{h}$, versus $\omega$ for $\tilde{q}_{s}=0.6,1.0,1.5$ with $r_{h}=1$, $\tilde{Q}=1$; non-positive everywhere, touching zero at $\omega=\tilde{q}_{s}\tilde{\phi}_{h}$.}
    \label{fig:wccc}
  \end{subfigure}\hfill
  \begin{subfigure}[t]{0.49\textwidth}
    \centering
    \includegraphics[width=\textwidth]{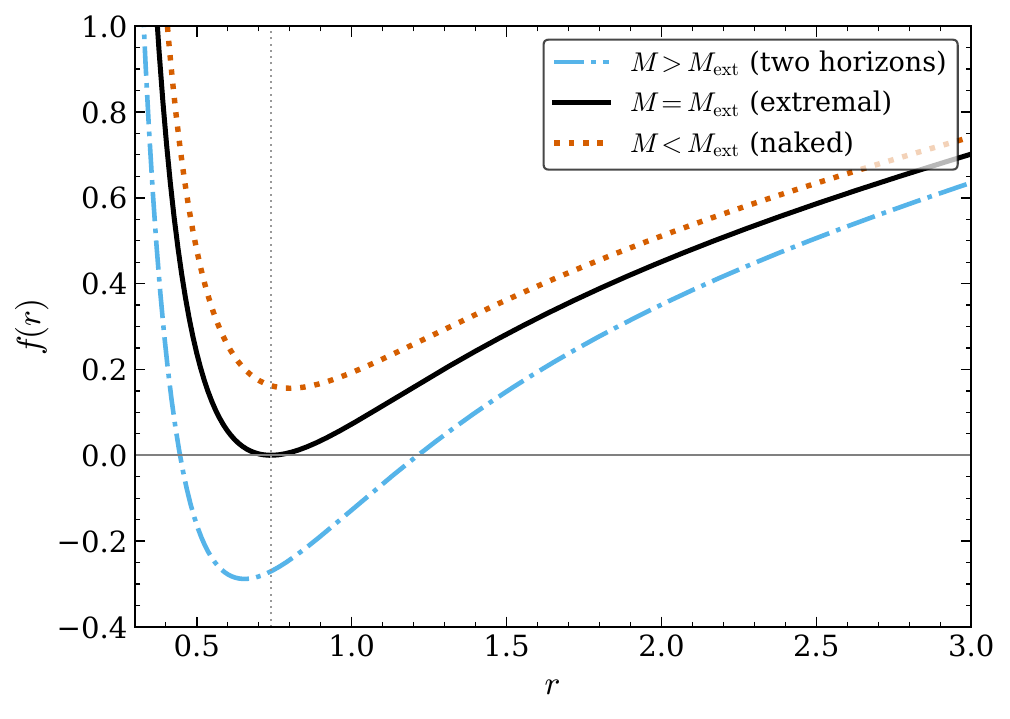}
    \caption{Metric function $f(r)$ near extremality at $Q=0.9$, $q=0.5$, $\gamma=1$, $\ell=8$: $M>M_{\mathrm{ext}}$ (two horizons), $M=M_{\mathrm{ext}}=0.75154$ (extremal, tangent at $r_{\mathrm{ext}}=0.73893$), and $M<M_{\mathrm{ext}}$ (naked).}
    \label{fig:nearext}
  \end{subfigure}
  \caption{Extremal censorship test (left) and the near-extremal horizon structure (right).}
  \label{fig:wccc_near}
\end{figure}

By varying the mass at fixed charge, Fig.~\ref{fig:nearext} sets out the three regimes. The extremal curve meets zero tangentially at $r_{\mathrm{ext}}$; the heavier configuration grows two horizons that bracket a negative minimum; the lighter one lifts the minimum above zero into the naked branch. What orders them is the sign of $f_{\min}$. Raising $M$ lowers the minimum through the $-2M/r$ term of Eq.~\eqref{eq:f}, lowering $M$ raises it, and the extremal mass is the boundary value where the minimum just grazes zero. That same control parameter runs the absorption process, since the flux $dM>0$ in Eq.~\eqref{eq:dMdQ2} drives the configuration deeper into the two-horizon region, not toward the naked branch.

That tangency in Fig.~\ref{fig:nearext} is the geometric content of the extremal condition $f(r_{\mathrm{ext}})=f'(r_{\mathrm{ext}})=0$. Since the curve meets the axis with zero slope, a small absorbed mass moves it down rather than up, keeping $f_{\min}\le0$ and protecting the horizon. The heavier curve crosses zero twice, which confirms the two-horizon count of Table~\ref{tab:horizon} for this parameter set; the lighter curve never reaches zero and would describe a naked singularity were it realized. Read with Eq.~\eqref{eq:ffinal_near}, the figure says that scalar absorption keeps the system on the safe side of the extremal boundary.

Evaluated at the outer horizon across $\gamma$, the derivatives of Eq.~\eqref{eq:partials} appear in Table~\ref{tab:dgamma}. The exponent derivative $\partial\Qcal/\partial\gamma$ is large and negative at small $\gamma$ and decays toward zero as $\gamma$ grows, a behavior set by the logarithmic and inverse-square factors of Eq.~\eqref{eq:dQdg}. The metric sensitivity $\partial f/\partial\gamma$ inherits that decay, since Eq.~\eqref{eq:partials} builds it from $\partial\Qcal/\partial\gamma$ and $\Qcal$, both falling with $\gamma$. The central-charge derivative $\partial f/\partial C$ stays small and negative throughout: it scales as $-r_{h}^{2}/(\ell^{2}C)$ at fixed large $\ell$, so a change in $C$ barely moves the minimum, and the conclusion of Sec.~\ref{isec4a} holds against modest $dC$.

\begin{table*}[htbp]
  \centering
  \setlength{\tabcolsep}{7pt}
  \renewcommand{\arraystretch}{1.25}
  \begin{tabularx}{\textwidth}{C C C C C C}
    \toprule
    \rowcolor{orange!50}
    $\gamma$ & $\Qcal$ & $\partial \Qcal/\partial \gamma$ & $r_{h}$ & $\partial f/\partial \gamma$ & $\partial f/\partial C$ \\
    \midrule
    0.90 & 0.81832 & $-5.47199$ & 2.00860 & $2.54064$ & $-0.0039400$ \\
    1.00 & 0.49000 & $-1.96990$ & 1.73400 & $1.01397$ & $-0.0029363$ \\
    1.30 & 0.22138 & $-0.40698$ & 1.49770 & $0.20995$ & $-0.0021906$ \\
    1.60 & 0.14238 & $-0.17038$ & 1.43010 & $0.07752$ & $-0.0019972$ \\
    2.00 & 0.09604 & $-0.07877$ & 1.39520 & $0.02803$ & $-0.0019008$ \\
    \bottomrule
  \end{tabularx}
  \caption{Sensitivity of the Yang--Mills amplitude and the metric function to the exponent $\gamma$ and the central charge $C$, evaluated at the outer horizon from Eqs.~\eqref{eq:partials} and~\eqref{eq:dQdg}, at $M=1$, $Q=0.9$, $q=0.7$, $\ell=8$ ($C=16$). The exponent sensitivity decays steeply with $\gamma$, while the central-charge sensitivity stays small and negative.}
  \label{tab:dgamma}
\end{table*}

\section{Mass--energy equivalence and the weak gravity conjecture}\label{isec5}

We now set $\hbar=c=1$, so that the energy of a scalar quantum equals its frequency, $E=\omega$, and its rest mass is $\mu_{s}$. Mass--energy equivalence then gives $\mu_{s}=\omega$, which lets us trade the frequency for the mass throughout. The horizon potential is $\tilde{\phi}_{h}=\tilde{Q}/r_{h}$, and the superradiance condition $\omega<\tilde{q}_{s}\tilde{\phi}_{h}$ becomes
\begin{equation}
\mu_{s}<\tilde{q}_{s}\tilde{\phi}_{h}\quad\Longleftrightarrow\quad\frac{\tilde{q}_{s}}{\mu_{s}}>\frac{1}{\tilde{\phi}_{h}}.
\label{eq:wgc1}
\end{equation}
The superradiant regime is therefore the regime in which the scalar satisfies a lower bound on its charge-to-mass ratio. For an extremal or near-extremal hole the horizon potential is largest, so the superradiance condition is also the condition for the field to obey the \WGC, up to the inverse-potential factor.

\subsection{The WGC threshold and its Yang--Mills modification}\label{isec5a}

To make the bound explicit we need $r_{\min}$. Differentiating Eq.~\eqref{eq:f},
\begin{equation}
f'(r)=\frac{2M}{r^{2}}+\frac{2r}{\ell^{2}}-\frac{2\tilde{Q}^{2}}{r^{3}}+\frac{(4\gamma-2)\Qcal}{r^{4\gamma-1}}.
\label{eq:fprime}
\end{equation}
Setting $f'(r_{\min})=0$ and multiplying by $r_{\min}^{3}$,
\begin{equation}
2Mr_{\min}+\frac{2r_{\min}^{4}}{\ell^{2}}-2\tilde{Q}^{2}+(4\gamma-2)\Qcal\,r_{\min}^{4-4\gamma}=0,
\label{eq:rmin_eq}
\end{equation}
which determines $r_{\min}$ implicitly. Solving for $M$,
\begin{equation}
M=\frac{\tilde{Q}^{2}}{r_{\min}}-\frac{r_{\min}^{3}}{\ell^{2}}-\frac{(4\gamma-2)\Qcal}{2}r_{\min}^{3-4\gamma}.
\label{eq:Mext}
\end{equation}
Substituting into the extremal condition $f(r_{\min})=0$ fixes the extremal radius $r_{\mathrm{ext}}$ for given $\tilde{Q}$, $\gamma$, $\ell$. Returning to the superradiance condition, the threshold is
\begin{equation}
\frac{\tilde{q}_{s}}{\mu_{s}}>\frac{1}{\tilde{\phi}_{h}}=\frac{r_{h}}{\tilde{Q}}.
\label{eq:wgc2}
\end{equation}
For a near-extremal hole, $r_{h}=r_{\min}+\varepsilon$, and expanding to first order,
\begin{equation}
\frac{1}{\tilde{\phi}_{h}}=\frac{r_{\min}}{\tilde{Q}}\left(1+\frac{\varepsilon}{r_{\min}}+O(\varepsilon^{2})\right),
\label{eq:invphi_exp}
\end{equation}
so the superradiant bound reads
\begin{equation}
\frac{\tilde{q}_{s}}{\mu_{s}}>\frac{r_{\min}}{\tilde{Q}}\left(1+\frac{\varepsilon}{r_{\min}}\right),
\label{eq:wgc_near}
\end{equation}
and the inequality reverses for absorption, $\omega>\tilde{q}_{s}\tilde{\phi}_{h}$. The standard \WGC\ asks for a state with $\tilde{q}_{s}/\mu_{s}\ge1$, but in AdS with a Yang--Mills sector the threshold is renormalized by the black-hole parameters. The quantity $r_{\min}/\tilde{Q}$ is the extremal inverse potential. For an extremal Reissner--Nordstr\"om hole without Yang--Mills one has $r_{\min}=\tilde{Q}$ and hence $r_{\min}/\tilde{Q}=1$, recovering the textbook bound. The Yang--Mills term shifts $r_{\min}$ relative to $\tilde{Q}$, so the effective bound becomes
\begin{equation}
\frac{\tilde{q}_{s}}{\mu_{s}}>\frac{r_{\min}}{\tilde{Q}}\quad(\text{extremal}),\qquad \frac{\tilde{q}_{s}}{\mu_{s}}>\frac{r_{\min}}{\tilde{Q}}\left(1+\frac{\varepsilon}{r_{\min}}\right)\quad(\text{near-extremal}).
\label{eq:wgc_final}
\end{equation}

\begin{figure}[htbp]
  \centering
  \begin{subfigure}[t]{0.49\textwidth}
    \centering
    \includegraphics[width=\textwidth]{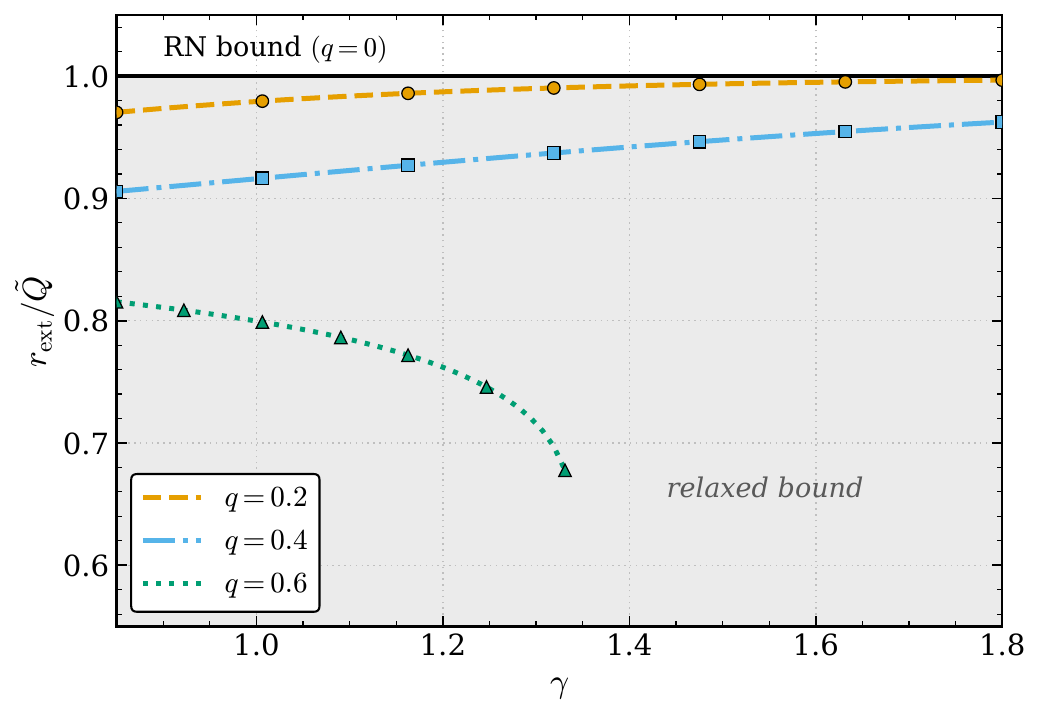}
    \caption{Effective \WGC\ threshold $r_{\mathrm{ext}}/\tilde{Q}$ versus $\gamma$ for $q=0.2,0.4,0.6$ at $\tilde{Q}=1$, $\ell=50$, against the Reissner--Nordstr\"om value of unity; the shaded band is the relaxed region, and the $q=0.6$ curve ends where the extremal configuration ceases to exist.}
    \label{fig:wgc}
  \end{subfigure}\hfill
  \begin{subfigure}[t]{0.49\textwidth}
    \centering
    \includegraphics[width=\textwidth]{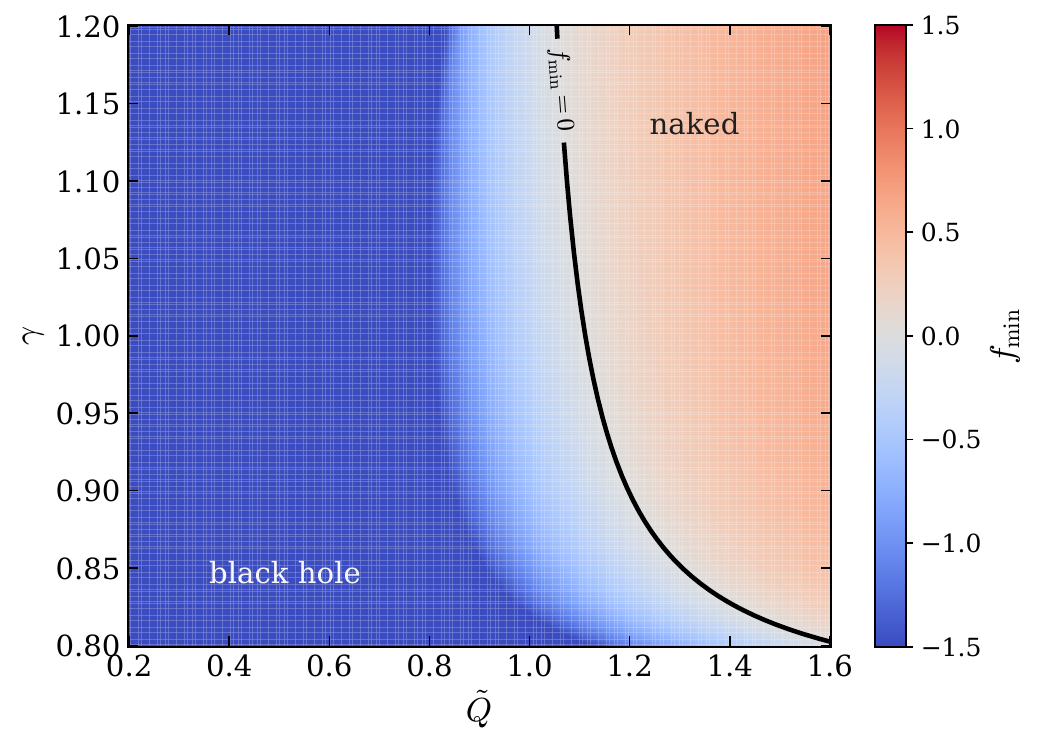}
    \caption{Censorship phase diagram: $f_{\min}$ over the $(\tilde{Q},\gamma)$ plane at $M=1$, $q=0.5$, $\ell=8$; the contour $f_{\min}=0$ separates the black-hole region (left) from the naked region (right).}
    \label{fig:phase}
  \end{subfigure}
  \caption{Effective \WGC\ threshold (left) and the censorship phase diagram (right).}
  \label{fig:wgc_phase}
\end{figure}

For three Yang--Mills charges, Fig.~\ref{fig:wgc} traces the threshold $r_{\mathrm{ext}}/\tilde{Q}$ across $\gamma$ against the Reissner--Nordstr\"om value of unity. Whenever $q>0$ the threshold sits below one, and it drops further as $q$ grows, so the non-Abelian sector relaxes the charge-to-mass bound a particle has to meet. The mechanism is the same negative Yang--Mills term of Eq.~\eqref{eq:f}: it pulls the extremal radius inward relative to the Maxwell case, which sends $r_{\mathrm{ext}}/\tilde{Q}$ below unity in Eq.~\eqref{eq:wgc_final}. The $q=0.6$ curve stops near $\gamma\simeq1.33$, where no extremal configuration survives for that charge, and this marks the edge of the window in which the bound is even defined. Set against the Reissner--Nordstr\"om--AdS analyses that pin the threshold at one~\cite{73}, the \EMPYM\ family admits lighter charged states and still protects the horizon.

At selected exponents the same threshold appears in Table~\ref{tab:wgc}. The $q=0$ column holds the Reissner--Nordstr\"om value to four digits for every $\gamma$, the consistency check that the Yang--Mills sector decouples once its charge vanishes. Raising $q$ lowers the threshold and shrinks the available range in $\gamma$, because the extremal condition~\eqref{eq:Mext} loses its solution when the Yang--Mills term overwhelms the Maxwell term.

\begin{table*}[htbp]
  \centering
  \setlength{\tabcolsep}{7pt}
  \renewcommand{\arraystretch}{1.25}
  \begin{tabularx}{\textwidth}{C C C C C}
    \toprule
    \rowcolor{orange!50}
    $\gamma$ & $q=0$ & $q=0.4$ & $q=0.7$ & $q=1.0$ \\
    \midrule
    0.90 & 0.99940 & 0.90916 & 0.74971 & 0.52702 \\
    1.00 & 0.99940 & 0.91605 & 0.71392 & -- \\
    1.30 & 0.99940 & 0.16723 & -- & -- \\
    1.60 & 0.99940 & 0.37308 & -- & -- \\
    2.00 & 0.99940 & 0.51353 & -- & -- \\
    \bottomrule
  \end{tabularx}
  \caption{Effective \WGC\ threshold $r_{\mathrm{ext}}/\tilde{Q}$ for several $\gamma$ and Yang--Mills charges $q$ at $\tilde{Q}=1$, $\ell=50$. The $q=0$ column reproduces the Reissner--Nordstr\"om bound of unity. A dash marks parameter pairs with no extremal configuration.}
  \label{tab:wgc}
\end{table*}

Because the censorship test of Sec.~\ref{isec4} and the \WGC\ threshold both turn on $r_{\min}$, they organize naturally into a single phase diagram. Figure~\ref{fig:phase} plots $f_{\min}$ over the $(\tilde{Q},\gamma)$ plane, with the contour $f_{\min}=0$ separating the black-hole region from the naked region. Small charge keeps $f_{\min}<0$ and a horizon exists, while large charge lifts $f_{\min}$ above zero and overcharges the configuration. The boundary curves upward in $\gamma$ because a larger exponent suppresses $\Qcal$ and lets the Maxwell charge reach overcharging at smaller $\tilde{Q}$. Absorption with $dM>0$ moves a configuration deeper into the black-hole region, consistent with Eq.~\eqref{eq:ffinal_ext3}, so the censorship boundary is never crossed by the scalar flux alone.

\subsection{Direction of evolution and the photon-sphere window}\label{isec5b}

The sign of $\Delta=\mu_{s}-\tilde{q}_{s}\tilde{\phi}_{h}$ fixes the direction of evolution. The mass and charge change by
\begin{equation}
dM=\mu_{s}(\mu_{s}-\tilde{q}_{s}\tilde{\phi}_{h})r_{h}^{2}\,dt,\qquad dQ=\tilde{q}_{s}(\mu_{s}-\tilde{q}_{s}\tilde{\phi}_{h})r_{h}^{2}\,dt,
\label{eq:dMdQ_wgc}
\end{equation}
and the leading shift of the near-extremal minimum is
\begin{equation}
f_{\mathrm{final}}=\delta-2(\mu_{s}-\tilde{q}_{s}\tilde{\phi}_{\min})^{2}r_{\min}\,dt+\frac{\partial f_{\min}}{\partial\gamma}d\gamma+\frac{\partial f_{\min}}{\partial C}dC+O(\varepsilon^{2}).
\label{eq:ffinal_wgc}
\end{equation}
The term $-2(\mu_{s}-\tilde{q}_{s}\tilde{\phi}_{\min})^{2}r_{\min}\,dt$ is non-positive and pushes $f_{\min}$ more negative, away from extremality. When $\mu_{s}>\tilde{q}_{s}\tilde{\phi}_{h}$ the hole absorbs energy and charge and moves toward extremality; when $\mu_{s}<\tilde{q}_{s}\tilde{\phi}_{h}$ it radiates them and recedes. A particle that satisfies the \WGC\ carries a high charge-to-mass ratio, so its emission drives the hole away from the extremal state, which is the direction that protects the horizon, in line with the connection drawn by Horowitz and Santos~\cite{Horowitz:2019eum}. Absorption of a sub-\WGC\ particle adds mass faster than charge and also recedes from extremality. In every case the horizon persists, so the \WGC\ selects the direction of motion in parameter space without forcing a censorship violation. This selector role parallels the charged-flux analysis of Anand and co-workers in perfect-fluid dark matter~\cite{Anand:2025wgc} and the Kerr--Newman--Kiselev--Letelier study of Noori Gashti and Pourhassan~\cite{Gashti:2025cqc}.

The threshold also has an optical face. The photon sphere of a static metric sits at the radius $r_{\mathrm{ps}}$ that solves $2f(r_{\mathrm{ps}})=r_{\mathrm{ps}}f'(r_{\mathrm{ps}})$. The same Yang--Mills term that shifts $r_{\min}$ in Eq.~\eqref{eq:rmin_eq} also shifts $r_{\mathrm{ps}}$, since it enters $f'$ in Eq.~\eqref{eq:fprime}. Written out for the \EMPYM\ metric, the condition is
\begin{equation}
2-\frac{6M}{r_{\mathrm{ps}}}+\frac{2Q^{2}}{r_{\mathrm{ps}}^{2}}-(4\gamma)\,\Qcal\,r_{\mathrm{ps}}^{2-4\gamma}=0,
\label{eq:photonsphere}
\end{equation}
where the AdS term drops out because $r^{2}/\ell^{2}$ satisfies $2f=rf'$ identically. A distant observer sees a shadow of impact parameter $b_{\mathrm{ps}}=r_{\mathrm{ps}}/\sqrt{f(r_{\mathrm{ps}})}$, which sets the angular size of the silhouette. Photon orbits are governed by the null effective potential
\begin{equation}
V_{\mathrm{ph}}(r)=\frac{f(r)}{r^{2}},
\label{eq:Vph}
\end{equation}
whose maximum sits at $r_{\mathrm{ps}}$. Here is the point. Because the extremal radius and the photon sphere both respond to $\gamma$ and $q$, a measured shadow size or lensing angle constrains the very parameters that fix the \WGC\ bound. The weight in Eq.~\eqref{eq:photonsphere} is $4\gamma$, so raising the exponent suppresses $\Qcal$ and lifts the power of $r_{\mathrm{ps}}^{2-4\gamma}$ at once, moving the photon sphere the same way as the extremal radius of Fig.~\ref{fig:wgc}. The link has been drawn before. For ModMax black holes the photon sphere tracks the \WGC\ threshold~\cite{Gashti:2025modmax}, and in gravity's rainbow the lensing observables encode the extremality bound~\cite{Gashti:2026lensing}. Thermodynamic-stability and shadow analyses of charged regular and AdS black holes point the same way~\cite{Sucu:2025crbh,Ahmed:2025bstr,50,8002,8003}. So the lowering of $r_{\mathrm{ext}}/\tilde{Q}$ in Fig.~\ref{fig:wgc} translates, through Eq.~\eqref{eq:photonsphere}, into a predicted shift of the photon-sphere radius that next-generation horizon-scale imaging could in principle bound.

\section{Numerical verification and parameter survey}\label{isec6}

The analytic results above were cross-checked with independent symbolic and numerical computation. Every equation of state was recomputed as a finite-difference derivative of $E$ and matched its closed form to better than $10^{-9}$ (Table~\ref{tab:eos_verify}), while the Euler relation and the conformal equation of state reproduced $E$ and $E/2$ to all digits (Table~\ref{tab:euler_verify}). Horizon structure, the extremal configuration, the metric minimum, and the \WGC\ threshold came from root finding on Eqs.~\eqref{eq:f}, \eqref{eq:fprime}, and~\eqref{eq:rmin_eq}, with the censorship boundary of Fig.~\ref{fig:phase} traced as the zero contour of $f_{\min}$.

The model also reduces correctly in its limits, a check on the construction itself. Table~\ref{tab:limits} gathers the metric value at a fixed reference radius across those limits. Three cases matter. As $q\to0$ the Yang--Mills amplitude $\Qcal$ vanishes and the solution becomes Reissner--Nordstr\"om--AdS, with $f$ matching the pure-charge value; at $\gamma\to1$ the power-Yang--Mills term returns to the standard Yang--Mills form and $\Qcal$ takes its $\gamma=1$ value; sending both the Yang--Mills and the charge term to zero leaves Schwarzschild--AdS. Each reduction works because the special case drops the corresponding term from Eq.~\eqref{eq:f} cleanly, with no residual dependence on the suppressed parameter. That is the behavior a consistent matter sector must show.

\begin{table*}[htbp]
  \centering
  \setlength{\tabcolsep}{6pt}
  \renewcommand{\arraystretch}{1.25}
  \begin{tabularx}{\textwidth}{L C C C C}
    \toprule
    \rowcolor{orange!50}
    limit & $q$ & $\gamma$ & $\Qcal$ & $f(r{=}2)$ \\
    \midrule
    $q\to0$ (RN--AdS)            & 0.0 & 1.30 & $1.41\times10^{-16}$ & 0.26500 \\
    $\gamma\to1$ (standard YM)   & 0.7 & 1.00 & 0.49000              & 0.14250 \\
    full \EMPYM                  & 0.7 & 1.30 & 0.22138              & 0.24091 \\
    Schwarzschild--AdS           & 0.0 & 1.00 & $1.00\times10^{-12}$ & 0.26500 \\
    \bottomrule
  \end{tabularx}
  \caption{Limiting behavior of the metric function at $r=2$, $M=1$, $Q=0.9$, $\ell=8$. The $q\to0$ and Schwarzschild--AdS rows return the pure-charge and vacuum values, the $\gamma\to1$ row recovers the standard Yang--Mills amplitude, and the full model interpolates between them.}
  \label{tab:limits}
\end{table*}

The Reissner--Nordstr\"om--AdS and Schwarzschild--AdS rows of Table~\ref{tab:limits} return the same $f(r{=}2)=0.265$, since both send the Yang--Mills amplitude to numerical zero and leave only the Maxwell and AdS terms, which coincide at this radius; the $\gamma=1$ row recovers the textbook amplitude $\Qcal=q^{2}$ for $n=2$, and the full model lands between. With the identity checks of Sec.~\ref{isec2d}, these limits confirm that the construction reduces to established results wherever a comparison exists.

\subsection{Comparison with related models}\label{isec6b}

It is useful to place these results next to earlier studies that tested the two conjectures together. For Reissner--Nordstr\"om--AdS black holes surrounded by perfect-fluid dark matter, Quyet found that the \WGC\ favors one branch of the CFT thermodynamics and that the censorship bound stays intact~\cite{Quyet:2026haf}; the \EMPYM\ family reaches the same compatibility, but through a non-Abelian sector rather than a dark-matter fluid, and with a threshold that moves below unity rather than staying pinned at it. Noori Gashti and Pourhassan reported a coinciding validity of both conjectures for the Kerr--Newman--Kiselev--Letelier black hole~\cite{Gashti:2025cqc}, where rotation and a Kiselev fluid play the role our Yang--Mills term plays here. Anand and collaborators used charged-scalar fluxes in the same CFT framework for charged AdS black holes in perfect-fluid dark matter~\cite{Anand:2025wgc}, and the flux structure of Eqs.~\eqref{eq:fluxes} matches theirs once the Yang--Mills amplitude is switched off.

The gedanken-experiment literature offers a second point of contact. He and Jiang showed that the \WCCC\ holds for Einstein--Born--Infeld black holes under second-order scalar absorption~\cite{He:2019dra}, and Jiang and Zhang reached the same conclusion for charged dilaton-Lifshitz black holes~\cite{Jiang:2020ald}. Our extremal result~\eqref{eq:ffinal_ext3}, with its non-positive perfect square, is the \EMPYM\ analogue of those findings, and it agrees with the general statement of Nat\'ario, Queimada, and Vicente that test fields cannot destroy extremal black holes~\cite{Natario:2016bay}. Anand, Mishra, and Channuie studied extremal stability and censorship in Kiselev spacetime~\cite{Anand:2025sek}, and Chen and collaborators turned censorship into constraints on low-energy effective theories~\cite{Chen:2020lbt}; both lines of work suggest that the lowered threshold of Fig.~\ref{fig:wgc} could feed into bounds on the allowed range of the nonlinearity exponent $\gamma$. The one qualitative difference from the Maxwell case is that the \EMPYM\ threshold is genuinely $\gamma$-dependent, so the model predicts a family of bounds rather than a single number, which is the feature that horizon-scale imaging~\cite{Gashti:2025modmax,Gashti:2026lensing} could in principle resolve.

\section{Island Formula and Information Recovery}\label{sec:island}

The information paradox starts from a simple mismatch. If backreaction is ignored, the
entanglement entropy of Hawking radiation grows without turning over, and that monotonic
growth breaks the unitarity of the quantum evolution. The island prescription removes the
mismatch by adding the non-perturbative gravitational contribution to the entanglement wedge
of the radiation region. With that contribution in place the radiation entropy follows the Page
curve: it rises to a maximum at the Page time $t_P$ and then settles to a constant.

We apply this prescription to the EMPYM--AdS black hole~\cite{Almheiri2021}. The aim is to
follow the entanglement entropy of the Hawking radiation in time and to show that the island
contribution restores the Page curve. The Page time turns out to be the useful object. It ties
the horizon geometry, the gauge charges, the Yang--Mills sector, and the AdS pressure into a
single number, so that information recovery in this model is set by the thermodynamic state of
the hole rather than by anything outside it.

\subsection{Setup and Kruskal coordinates}\label{isec:island_setup}

We work with the EMPYM--AdS geometry of Section~\ref{isec2}. The lapse function $f(r)$, the
Yang--Mills sector, and the bulk mass parameter are the ones defined there. In four dimensions
($n=2$) with $G=1$, the surface gravity at the event horizon is
\begin{equation}
\kappa=\frac{f'(r_h)}{2}=\frac{1}{2}\left[\frac{2M}{r_h^{2}}+\frac{2r_h}{\ell^{2}}
-\frac{2Q^{2}}{r_h^{3}}+\frac{(4\gamma-2)\,\Qcal}{r_h^{4\gamma-1}}\right],
\label{eq:kappa}
\end{equation}
and the Hawking temperature follows as $T=\kappa/(2\pi)$.

To track the entanglement structure of the radiation we pass to Kruskal coordinates,
\begin{equation}
U=-e^{-\kappa(t-r_*)},\qquad V=e^{\kappa(t+r_*)},\qquad r_*=\int\frac{dr}{f(r)},
\label{eq:kruskal}
\end{equation}
in which the metric is conformally flat, $ds^{2}=W(r)^{2}\,dU\,dV$, with
\begin{equation}
W(r)^{2}=\frac{f(r)}{\kappa^{2}e^{2\kappa r_*}}.
\label{eq:Wconf}
\end{equation}
The radiation region is $R=\,]-\infty,b_-]\cup[b_+,+\infty[$, with boundary points at
$(t_b,b)$ for $b_+$ and $(-t_b+i\beta/2,b)$ for $b_-$, where $\beta=1/T$.

\subsection{Entanglement entropy without island}\label{isec:island_noisland}

Before the Page time there is no island, and the entropy comes from the standard CFT
two-point function in the Kruskal background~\cite{Hashimoto2020},
\begin{equation}
S(R)=\frac{c}{3}\log\!\left(\frac{2\cosh(\kappa t_b)}{\kappa}\right)+\frac{c}{6}\log f(b),
\label{eq:Snoisland}
\end{equation}
with $c$ the central charge of the dual CFT and $f(b)$ the metric function~\eqref{eq:f}
evaluated at the detector position $b$. Written out,
\begin{equation}
\frac{c}{6}\log f(b)=\frac{c}{6}\log\!\left(1-\frac{2M}{b}+\frac{b^{2}}{\ell^{2}}
+\frac{Q^{2}}{b^{2}}-\frac{\Qcal}{b^{4\gamma-2}}\right).
\label{eq:logfb}
\end{equation}
At late times the cosh term dominates. For $t_b\to\infty$ Eq.~\eqref{eq:Snoisland} reduces to
\begin{equation}
S(R)\approx\frac{c}{3}\kappa\,t_b=\frac{c}{2}\,T\,t_b,
\label{eq:Slinear}
\end{equation}
a linear, unbounded growth. This is exactly the behaviour that breaks unitarity, and it is what
the island contribution has to cure.

\subsection{Entanglement entropy with island}\label{isec:island_island}

Past the Page time an island region $I$ forms inside the hole, with boundaries $a_\pm$ at
$(t_a,a)$ and $(-t_a+i\beta/2,a)$. The generalised entropy is~\cite{Almheiri2021}
\begin{equation}
S_{\rm gen}(R)=2\pi a^{2}+\frac{c}{3}\log\!
\frac{d(a_+,a_-)\,d(b_+,b_-)\,d(a_+,b_+)\,d(a_-,b_-)}{d(a_+,b_-)\,d(a_-,b_+)},
\label{eq:Sgen}
\end{equation}
with the geodesic distances taken in the Kruskal frame. In the conformally flat sector the
interval is $ds^{2}=W(r)^{2}\,dU\,dV$, so near the horizon
\begin{equation}
d(x,y)=\sqrt{(U_x-U_y)(V_x-V_y)},
\label{eq:geodist}
\end{equation}
valid up to an overall conformal factor that drops out of the entropy
ratios~\cite{Almheiri2021,Hashimoto2020}. Extremising in the island time, $\partial S_{\rm
gen}/\partial t_a=0$, fixes $t_a=t_b$, and at late times the distance hierarchy
\begin{equation}
d(a_+,a_-)\simeq d(b_+,b_-)\simeq d(a_+,b_-)\simeq d(a_-,b_+)\gg d(a_+,b_+)\simeq d(a_-,b_-)
\label{eq:hierarchy}
\end{equation}
collapses Eq.~\eqref{eq:Sgen} to
\begin{equation}
S_{\rm gen}(R)=2\pi a^{2}+\frac{c}{3}\log\!\big[d(a_+,b_+)\,d(a_-,b_-)\big].
\label{eq:Sgen_simpl}
\end{equation}
The remaining condition $\partial S_{\rm gen}/\partial a=0$ puts the island boundary just
outside the horizon, $a=r_h+\chi^{2}r_h$ with $\chi\ll1$. With the near-horizon forms $f(a)\approx
2\kappa r_h\chi^{2}$ and $r_*(a)\approx\kappa^{-1}\log\chi$ one finds
\begin{equation}
\chi=\frac{c\,e^{-\kappa r_*(b)}}{12\pi r_h^{2}},
\label{eq:chi}
\end{equation}
so the island surface sits at
\begin{equation}
a=r_h+\left(\frac{c\,e^{-\kappa r_*(b)}}{12\pi r_h^{2}}\right)^{2}r_h.
\label{eq:island_loc}
\end{equation}
Keeping the leading term in Eq.~\eqref{eq:Sgen_simpl} gives
\begin{equation}
S(R)\approx 2S=2\pi r_h^{2}.
\label{eq:Ssat}
\end{equation}
The entropy locks to twice the Bekenstein--Hawking value. The Page curve is recovered, and
unitarity with it.

\subsection{Page time}\label{isec:island_pagetime}

The Page time is set by matching the two phases, the linear branch~\eqref{eq:Slinear} against
the saturated value~\eqref{eq:Ssat},
\begin{equation}
\frac{c}{2}\,T\,t_P=2\pi r_h^{2},
\label{eq:pagematch}
\end{equation}
which gives
\begin{equation}
t_P=\frac{4\pi r_h^{2}}{c\,T}=\frac{16\pi^{2}r_h^{2}}{c\,f'(r_h)}.
\label{eq:tP}
\end{equation}
Inserting the surface gravity~\eqref{eq:kappa} of the EMPYM--AdS black hole,
\begin{equation}
t_P=\frac{16\pi^{2}r_h^{2}}
{c\left[\dfrac{2M}{r_h^{2}}+\dfrac{2r_h}{\ell^{2}}-\dfrac{2Q^{2}}{r_h^{3}}
+\dfrac{(4\gamma-2)\,\Qcal}{r_h^{4\gamma-1}}\right]},
\label{eq:tP_explicit}
\end{equation}
and, trading the AdS term for the pressure through $P=3/(8\pi\ell^{2})$, i.e.
$2r_h/\ell^{2}=16\pi P r_h/3$,
\begin{equation}
t_P=\frac{16\pi^{2}r_h^{2}}
{c\left[\dfrac{2M}{r_h^{2}}+\dfrac{16\pi P r_h}{3}-\dfrac{2Q^{2}}{r_h^{3}}
+\dfrac{(4\gamma-2)\,\Qcal}{r_h^{4\gamma-1}}\right]}.
\label{eq:tP_pressure}
\end{equation}
Equation~\eqref{eq:tP_explicit} carries the full thermodynamic state of the hole, through
$r_h$, $Q$, $\Qcal$, $\gamma$, and $P$.

\begin{figure}[htbp]
  \centering
  \includegraphics[width=0.98\textwidth]{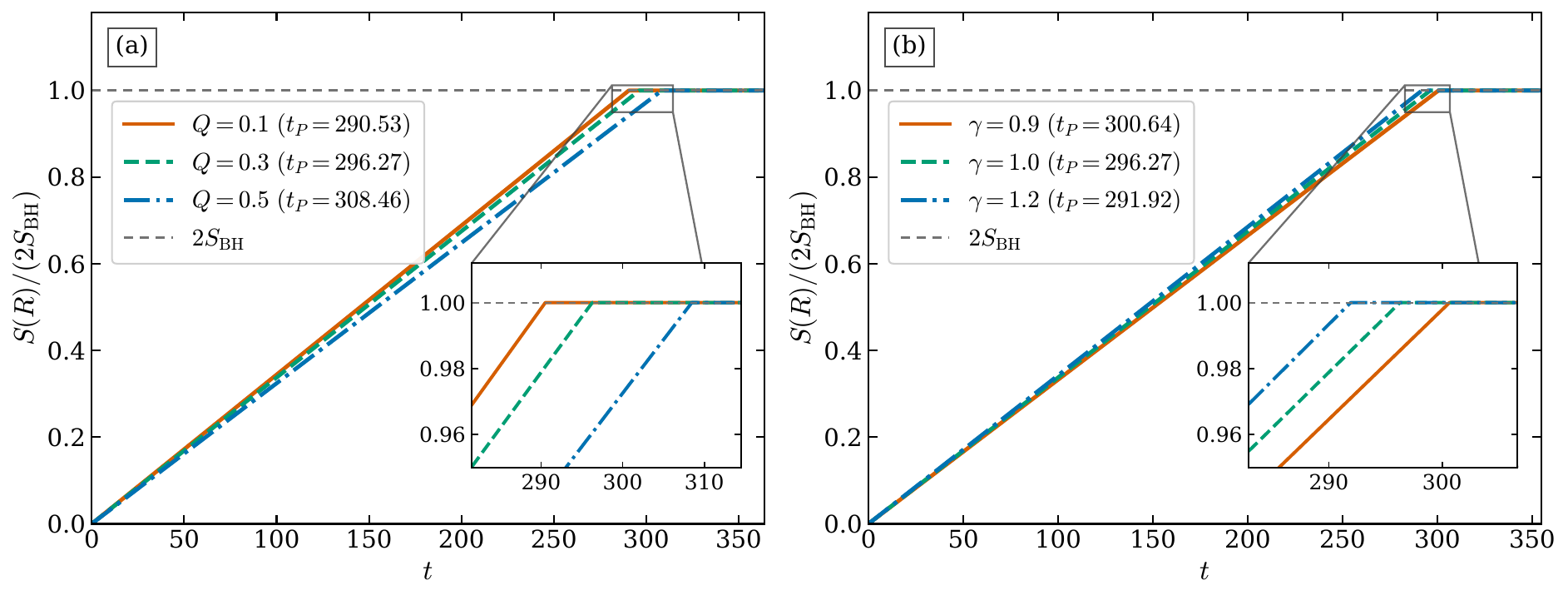}
  \caption{Normalised entanglement entropy $S(R)/(2S_{\rm BH})$ as a function of time $t$ for
  EMPYM--AdS black holes at $r_h=1.5$, $\Qcal=0.2$, $\ell=3$, $c=1$. (a) Maxwell charge
  $Q=0.1,0.3,0.5$ at $\gamma=1$. (b) Yang--Mills exponent $\gamma=0.9,1.0,1.2$ at $Q=0.3$. Each
  curve rises linearly while Hawking radiation dominates and then saturates at $S=2S_{\rm BH}$
  once the island forms. The ceiling $2\pi r_h^{2}$ is fixed by $r_h$, so the curves share a
  common saturation level but reach it at different Page times; the legend lists $t_P$ for each
  case. A larger $Q$ lowers the temperature and flattens the early slope, moving the turnover to
  later $t$, while a larger $\gamma$ steepens the slope and brings it forward. The inset
  magnifies the saturation corners, where the three Page times separate. The dashed line marks
  $2S_{\rm BH}$.}
  \label{fig:pagecurve}
\end{figure}

Figure~\ref{fig:pagecurve} shows the time evolution of the normalised entanglement entropy for
several Maxwell charges $Q$ and Yang--Mills exponents $\gamma$. Every curve traces the same
Page shape: a linear Hawking-dominated rise, then saturation at $S(R)/(2S_{\rm BH})=1$ once the
island switches on. What $Q$ and $\gamma$ change is not the shape but the effective temperature,
and through it the slope and the Page time. A larger $Q$ lowers the temperature and flattens the
early growth; a larger $\gamma$ steepens it. The ceiling never moves. Fixed by $r_h$ alone, it
forces the curves to share a saturation level and to separate only in where they turn over,
which we read as information recovery governed by thermodynamic scaling rather than by the
microscopic detail of the gauge sector.

Figure~\ref{fig:tP_rh} follows the Page time against the horizon radius. It is sensitive to the
state variables. Raising $Q$ pushes the island phase later by increasing $t_P$, while raising
$\gamma$ or the pressure $P$ brings it forward. So the same thermodynamic data that fixes the
phase structure also sets the clock for information recovery.

\subsection{Page curve and the thermodynamic--information correspondence}\label{isec:island_corr}

The two phases combine into one picture. Without an island the entanglement entropy grows as
$S(R)\propto T t_b$ and runs off at late times, in conflict with unitarity. With the island it
saturates at $2\pi r_h^{2}$, and the Page curve returns. The Page time~\eqref{eq:tP_explicit} is
the bridge between the two, and reading off its terms shows how each piece of the
thermodynamics acts.

\begin{figure}[htbp]
  \centering
  \includegraphics[width=0.98\textwidth]{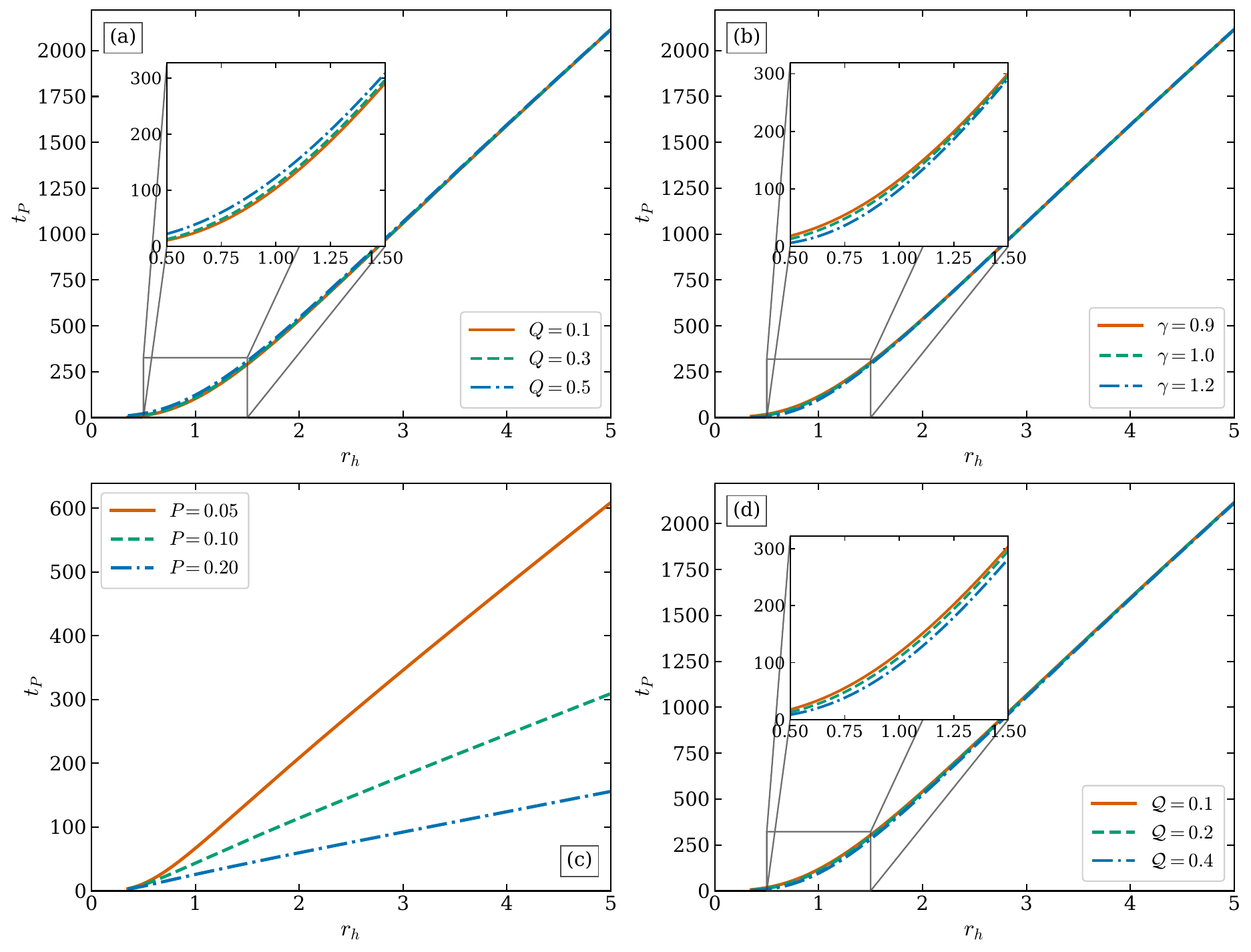}
  \caption{Page time $t_P$ versus event-horizon radius $r_h$ for EMPYM--AdS black holes ($c=1$).
  (a) $Q=0.1,0.3,0.5$ at $\Qcal=0.2$, $\gamma=1$, $\ell=3$. (b) $\gamma=0.9,1.0,1.2$ at
  $Q=0.3$, $\Qcal=0.2$, $\ell=3$. (c) $P=0.05,0.10,0.20$ at $Q=0.3$, $\Qcal=0.2$,
  $\gamma=1$. (d) $\Qcal=0.1,0.2,0.4$ at $Q=0.3$, $\gamma=1$, $\ell=3$. A larger $Q$ raises
  $t_P$ and slows recovery; a larger $\gamma$ or a higher $P$ lowers it. In panels (a), (b), and
  (d) the three curves nearly coincide and separate only at small $r_h$, where the gauge terms
  compete with the AdS term, so the insets magnify the window $r_h\in[0.5,1.5]$ to resolve that
  splitting. Pressure (panel c) separates the curves over the whole range, and needs no
  inset~\cite{Ladghami2026}.}
  \label{fig:tP_rh}
\end{figure}

The $-2Q^{2}/r_h^{3}$ term lowers the denominator and therefore lifts $t_P$. A larger Maxwell
charge delays the island phase and slows recovery, which fits the lower Hawking temperature of a
charged hole. The Yang--Mills term $(4\gamma-2)\Qcal/r_h^{4\gamma-1}$ is positive for
$\gamma>1/2$; it enlarges the denominator and shortens $t_P$, so a larger $\gamma$ speeds
recovery, most strongly for small holes. From Eq.~\eqref{eq:tP_pressure} a higher pressure
shortens $t_P$ at every radius, matching what is seen for Schwarzschild--AdS black holes in
conformal Killing gravity~\cite{Ladghami2026}. Small holes are hot and evaporate quickly, so
$t_P$ is short there. As $r_h$ grows the competition among the terms in
Eq.~\eqref{eq:tP_explicit} decides whether $t_P$ rises or falls, which is the origin of the
two-regime behaviour in Fig.~\ref{fig:tP_rh}.

\subsection{Critical behaviour of the Page time}\label{isec:island_critical}

The critical behaviour of AdS black holes has been studied in detail within the extended
phase-space formalism~\cite{Kubiznak2012,Gunasekaran2012,Wei2013,Cai2013}, and the link between
that criticality and information recovery has been examined for several
systems~\cite{Kubiznak2017,Lin2024,Yu2022,Ladghami2026}. We carry the analysis over to the
EMPYM--AdS case. Using Eq.~\eqref{eq:kappa} and $P=3/(8\pi\ell^{2})$, the equation of state is
\begin{equation}
P=\frac{T}{2r_+}-\frac{1}{8\pi r_+^{2}}+\frac{Q^{2}}{8\pi r_+^{4}}
-\frac{(4\gamma-2)\,\Qcal}{8\pi\,r_+^{4\gamma}}.
\label{eq:eos}
\end{equation}
The critical point follows from
\begin{equation}
\left(\frac{\partial P}{\partial r_+}\right)_{T}=0,\qquad
\left(\frac{\partial^{2}P}{\partial r_+^{2}}\right)_{T}=0.
\label{eq:critcond}
\end{equation}
The first condition gives
\begin{equation}
T_c=\frac{1}{2\pi r_c^{2}}-\frac{Q^{2}}{\pi r_c^{4}}
+\frac{(4\gamma-2)\cdot4\gamma\,\Qcal}{4\pi\,r_c^{4\gamma}},
\label{eq:Tc}
\end{equation}
and substituting it into the second yields
\begin{equation}
\frac{T_c}{r_c^{3}}-\frac{3}{4\pi r_c^{4}}+\frac{5Q^{2}}{2\pi r_c^{6}}
-\frac{(4\gamma-2)\cdot4\gamma\cdot(4\gamma+1)\,\Qcal}{8\pi\,r_c^{4\gamma+2}}=0,
\label{eq:rc}
\end{equation}
solved numerically for given $Q$, $\Qcal$, $\gamma$. The critical pressure $P_c$ then
follows from Eq.~\eqref{eq:eos}, and the critical Page time is
\begin{equation}
t_c=\frac{4\pi r_c^{2}}{c\,T_c}.
\label{eq:tc}
\end{equation}
With the reduced variables
\begin{equation}
p=\frac{P}{P_c},\qquad \rho=\frac{r_+}{r_c},\qquad \tau_P=\frac{t_P}{t_c},
\label{eq:reduced}
\end{equation}
the normalised Page time reads
\begin{equation}
\tau_P(\rho,p)=\frac{r_+^{2}\,T_c}{r_c^{2}\,T(r_+,pP_c)}.
\label{eq:tauP}
\end{equation}
The critical values are listed in Table~\ref{tab:critical}.

\begin{figure}[htbp]
  \centering
  \includegraphics[width=0.98\textwidth]{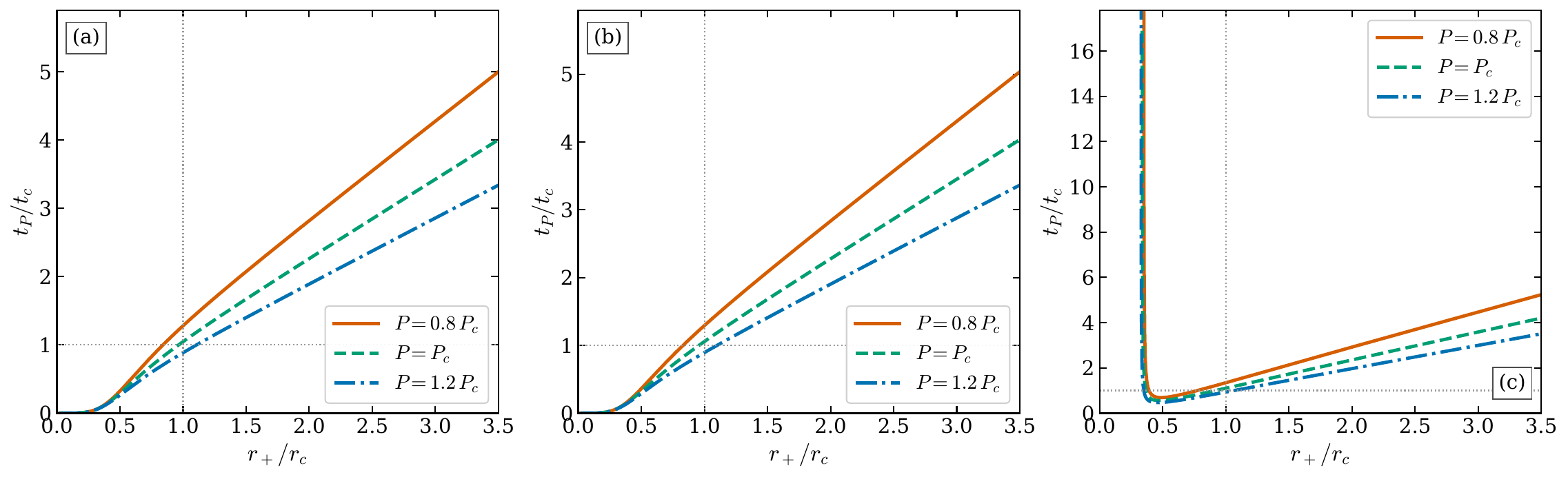}
  \caption{Normalised Page time $t_P/t_c$ versus $r_+/r_c$ at $\Qcal=0.2$, $\gamma=1$,
  $c=1$, for (a) $Q=0.1$, (b) $Q=0.3$, (c) $Q=0.5$. Red, green, and blue denote $P=0.8P_c$,
  $P_c$, and $1.2P_c$; the dotted lines mark $r_+/r_c=1$ and $t_P/t_c=1$. A higher pressure
  lowers $t_P/t_c$ throughout. In the large-charge panel (c) the effective temperature passes
  through zero near $r_+/r_c\simeq0.4$: the hole reaches a cold, near-extremal state and $t_P$
  diverges, which produces the spike that sets (c) apart from the smooth small-charge cases (a)
  and (b)~\cite{Kubiznak2012,Wei2013}.}
  \label{fig:tP_varyQ}
\end{figure}

\begin{figure}[htbp]
  \centering
  \includegraphics[width=0.98\textwidth]{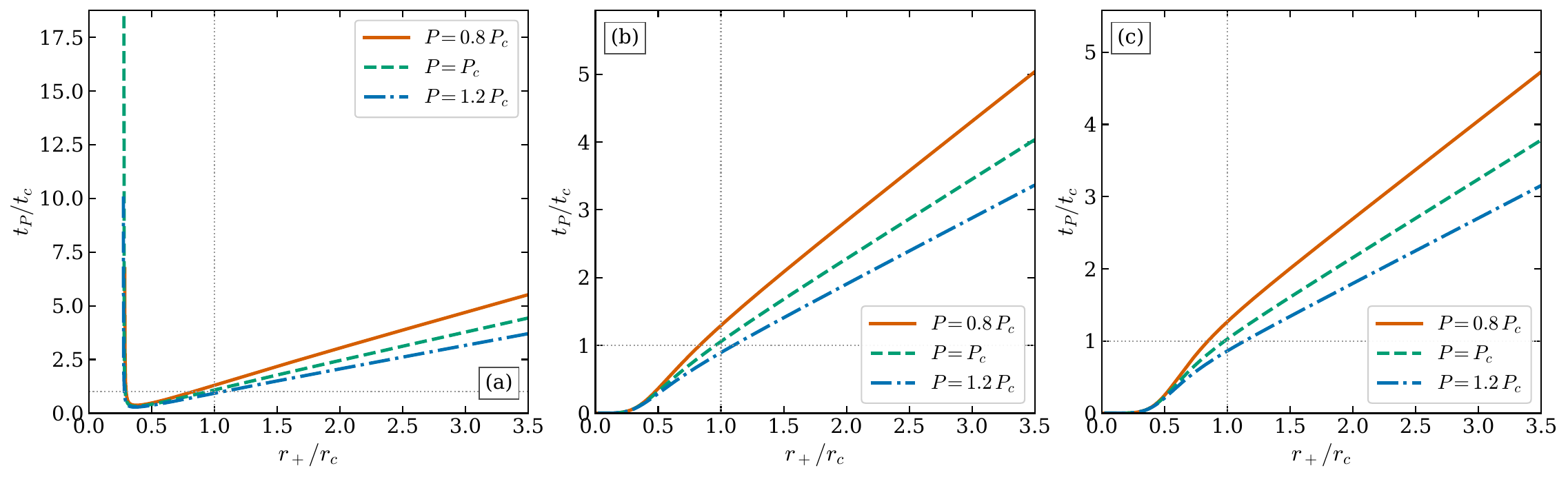}
  \caption{Normalised Page time $t_P/t_c$ versus $r_+/r_c$ at $Q=0.3$, $\Qcal=0.2$, $c=1$,
  for (a) $\gamma=0.9$, (b) $\gamma=1.0$, (c) $\gamma=1.2$. Colour coding as in
  Fig.~\ref{fig:tP_varyQ}. The spike in panel (a) marks the radius where $T\to0$: at $\gamma=0.9$
  the Maxwell term controls the small-$r_+$ temperature, drives it through zero, and $t_P$
  diverges there. For $\gamma\ge1$ the Yang--Mills term keeps the temperature positive and the
  curves stay smooth. A larger $\gamma$ also lowers $r_c$ and $t_c$ while raising $T_c$ and $P_c$
  (Table~\ref{tab:critical}), yet the smooth panels keep the same shape, so the pressure
  dependence of $\tau_P$ is set by the critical point rather than by $\gamma$
  directly~\cite{Lin2024}.}
  \label{fig:tP_varyGamma}
\end{figure}

\begin{figure}[htbp]
  \centering
  \includegraphics[width=0.98\textwidth]{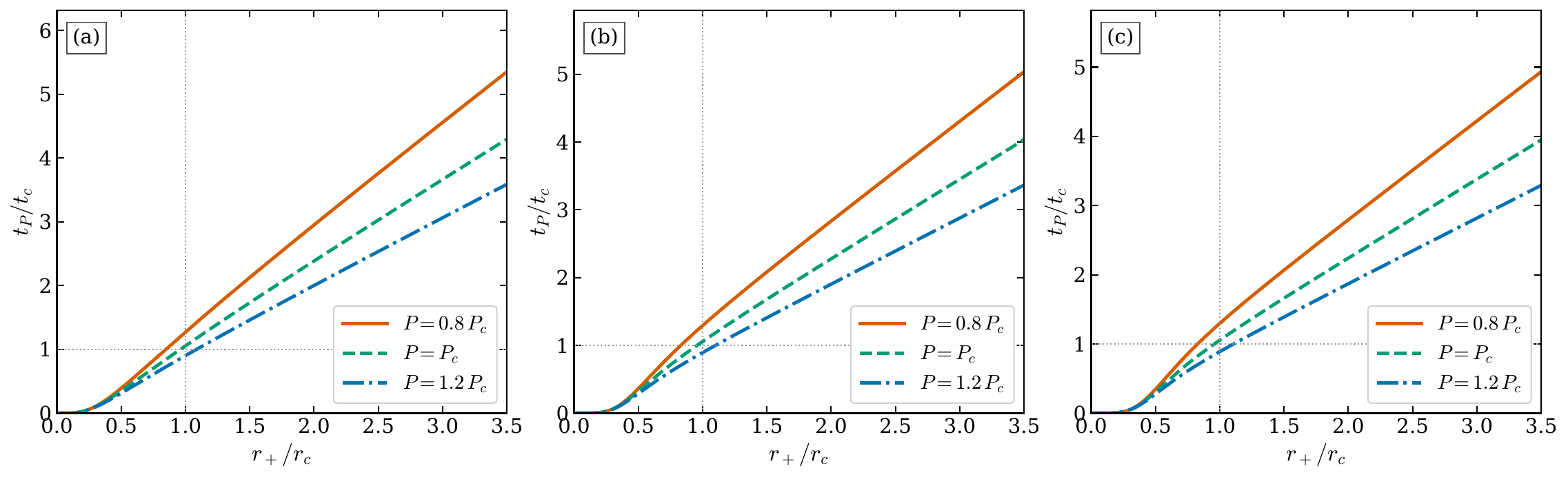}
  \caption{Normalised Page time $t_P/t_c$ versus $r_+/r_c$ at $Q=0.3$, $\gamma=1$, $c=1$, for (a)
  $\Qcal=0.1$, (b) $\Qcal=0.2$, (c) $\Qcal=0.4$. Colour coding as in
  Fig.~\ref{fig:tP_varyQ}. The temperature stays positive for all three amplitudes, so every
  panel is smooth. A larger $\Qcal$ raises $T_c$ and $P_c$ and reduces $r_c$ and $t_c$,
  which moves the knee of each curve slightly inward; the panels differ in that knee position
  while sharing the same pressure ordering.}
  \label{fig:tP_varyQcal}
\end{figure}

\begin{figure}[htbp]
  \centering
  \includegraphics[width=0.92\textwidth]{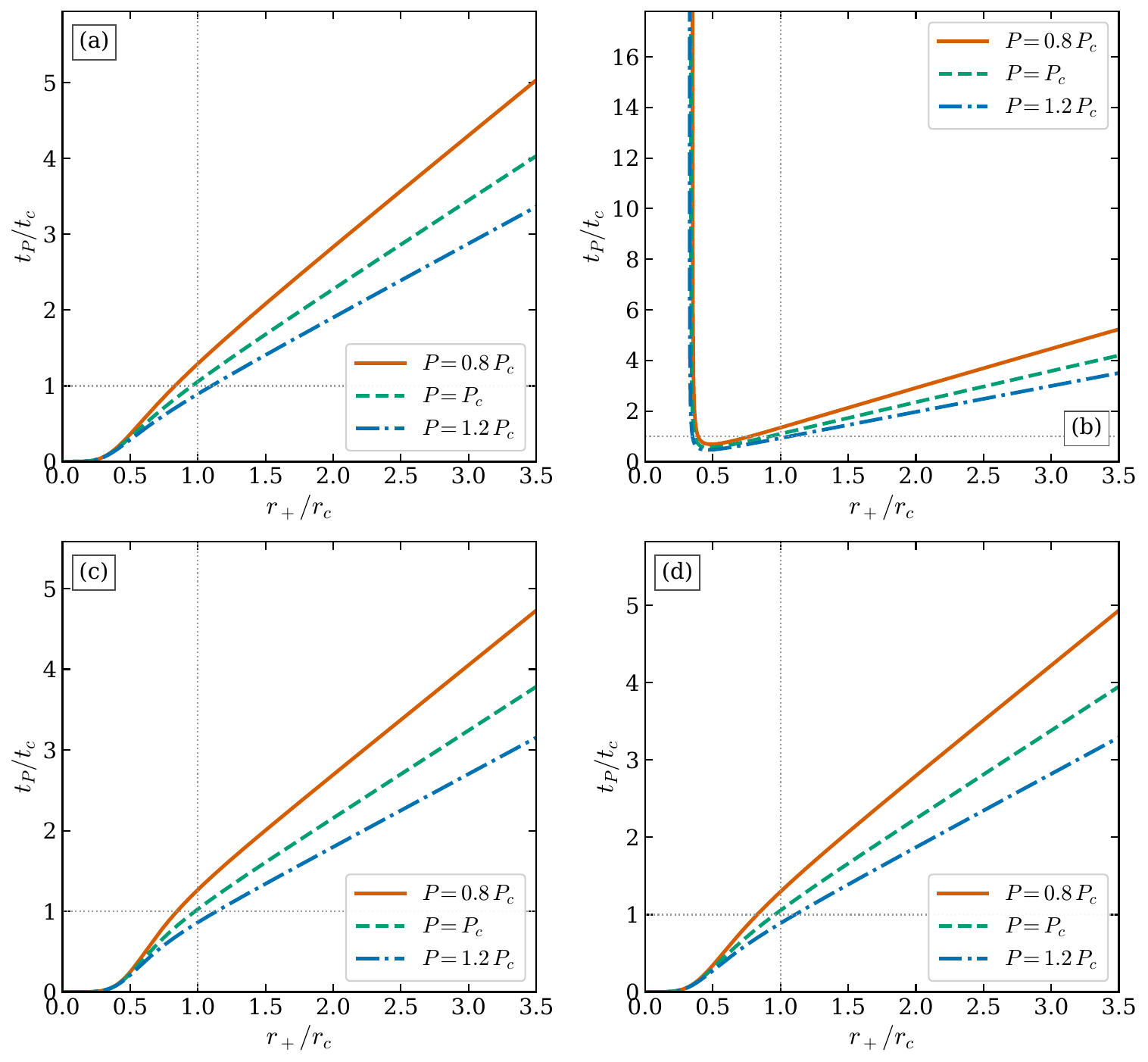}
  \caption{Summary of $t_P/t_c$ versus $r_+/r_c$ for four EMPYM--AdS parameter sets ($c=1$): (a)
  $Q=0.3$, $\Qcal=0.2$, $\gamma=1$ (reference); (b) $Q=0.5$, $\Qcal=0.2$, $\gamma=1$;
  (c) $Q=0.3$, $\Qcal=0.2$, $\gamma=1.2$; (d) $Q=0.3$, $\Qcal=0.4$, $\gamma=1$. Red,
  green, blue: $P=0.8P_c$, $P_c$, $1.2P_c$. A higher pressure lowers $\tau_P$ in every panel.
  Only the large-Maxwell-charge case (b) develops the near-extremal spike where $T\to0$; the
  other three stay smooth. The figure therefore singles out the Maxwell charge as the parameter
  that drives the system toward the cold, information-trapping
  limit~\cite{Almheiri2021,Kubiznak2012}.}
  \label{fig:tP_summary}
\end{figure}

The ratio $t_P/t_c$ in Figs.~\ref{fig:tP_varyQ}--\ref{fig:tP_summary} gives a scale-free probe
of information recovery. Rescaling by the critical point removes the absolute thermodynamic
scale and partly factors out the parameter dependence, which lets the shared structure show
through. Across the panels the curves vary monotonically with $\rho=r_+/r_c$ and turn over near
$\rho=1$. Their detailed shape still depends on $Q$, $\gamma$, and $\Qcal$.

Three trends stand out. First, raising the Maxwell charge $Q$ lifts the curves, a global delay
of information recovery across scales. Second, raising the Yang--Mills exponent $\gamma$ deforms
the curve most strongly at small $r_+$, where the nonlinear gauge term is largest and the
curvature is high. Third, $\Qcal$ acts mainly in the intermediate range, where the Maxwell
and Yang--Mills terms are comparable. There is also a sharper effect worth naming. For the
large-charge case $Q=0.5$ (Fig.~\ref{fig:tP_varyQ}c, Fig.~\ref{fig:tP_summary}b) and for the
small exponent $\gamma=0.9$ (Fig.~\ref{fig:tP_varyGamma}a), the effective temperature passes
through zero at a finite $\rho$; the hole approaches a cold, near-extremal state and $t_P$
diverges, which shows up as the spike in those panels.

We first read this family of curves as a clean collapse onto a single profile. It is not quite
that. The small-$r_+$ shapes still separate by parameter, and the spiked panels stand apart from
the smooth ones, so what survives the rescaling is a near-critical scaling near $\rho=1$ rather
than a full collapse. In that sense the Page time behaves as a critical observable in the
extended phase space: it inherits the thermodynamic phase structure of the black hole, while the
gauge content controls how sharply the cold-hole limit is approached.

\begin{table}[htbp]
\centering
\renewcommand{\tabcolsep}{12pt}
\renewcommand{\arraystretch}{1.6}
\caption{Critical thermodynamic quantities and critical Page time for EMPYM--AdS black holes
($c=G=1$), obtained numerically from Eqs.~\eqref{eq:Tc} and~\eqref{eq:rc}. The three groups show
the separate effect of varying $Q$, $\gamma$, and $\Qcal$.}
\label{tab:critical}
\begin{tabular}{ccccccc}
\hline\hline
\rowcolor{orange!50}
$Q$ & $\Qcal$ & $\gamma$ & $r_c$ & $T_c$ & $P_c$ & $t_c$ \\
\hline
0.1 & 0.2 & 1.0 & 0.4337 & 4.3546 & 4.3702 & 0.5428 \\
0.3 & 0.2 & 1.0 & 0.4425 & 3.3871 & 3.3025 & 0.7264 \\
0.5 & 0.2 & 1.0 & 0.4854 & 1.5354 & 1.3052 & 1.9285 \\
0.3 & 0.2 & 0.9 & 0.5503 & 1.0005 & 0.7074 & 3.8034 \\
0.3 & 0.2 & 1.0 & 0.4425 & 3.3871 & 3.3025 & 0.7264 \\
0.3 & 0.2 & 1.2 & 0.3487 & 32.988 & 43.717 & 0.0463 \\
0.3 & 0.1 & 1.0 & 0.5109 & 1.1237 & 0.8831 & 2.9187 \\
0.3 & 0.2 & 1.0 & 0.4425 & 3.3871 & 3.3025 & 0.7264 \\
0.3 & 0.4 & 1.0 & 0.4184 & 8.2865 & 8.7540 & 0.2654 \\
\hline\hline
\end{tabular}
\end{table}

\section{Conclusions}\label{isec7}

We studied the weak cosmic censorship and weak gravity conjectures for four-dimensional \EMPYM\ black holes within holographic CFT thermodynamics, using a charged massive scalar field as the perturbing agent. Starting from the metric function and the holographic dictionary, we built the equations of state, derived and verified an Euler relation, computed the horizon fluxes, followed the metric minimum through absorption, and connected the resulting \WGC\ threshold to the photon sphere. The thermodynamic foundation rests on the extended first law $dE=\tilde{T}\,dS+\tilde{\phi}\,d\tilde{Q}+\tilde{\psi}\,d\tilde{q}+\mu\,dC-p\,d\Vol$, in which the central charge $C$ and the volume $\Vol$ enter as independent variables and the Yang--Mills potential $\tilde{\psi}$ together with the chemical potential $\mu$ carry the exponent $\gamma$ explicitly through Eqs.~\eqref{eq:psi} and~\eqref{eq:mu}. Because the internal energy is homogeneous in its arguments, it also obeys the Euler relation $E=\tilde{T}S+\tilde{\phi}\tilde{Q}+\tfrac{3\gamma-1}{2\gamma}\tilde{\psi}\tilde{q}+\mu C$ and the conformal equation of state $p\,\Vol=E/2$; both were confirmed to machine precision, which fixes the thermodynamics with no free integration constant.

The charged scalar then supplies a controlled perturbation. Its horizon fluxes are $dE/dt=\omega(\omega-\tilde{q}_{s}\tilde{\phi}_{h})r_{h}^{2}$ and $dQ/dt=\tilde{q}_{s}(\omega-\tilde{q}_{s}\tilde{\phi}_{h})r_{h}^{2}$, so the field deposits energy and charge for $\omega>\tilde{q}_{s}\tilde{\phi}_{h}$ and extracts them in the superradiant regime $\omega<\tilde{q}_{s}\tilde{\phi}_{h}$. Following the minimum of $f(r)$ through the absorption settles the censorship question. For an extremal hole the leading change of the minimum reduces to a perfect square, $-2(\omega-\tilde{q}_{s}\tilde{\phi}_{h})^{2}r_{h}\,dt\le0$, so with the couplings held fixed the minimum stays non-positive and the horizon survives; the near-extremal case gives the same answer to leading order in the small parameter $\varepsilon$. The Yang--Mills sector changes the quantitative picture without overturning it. The mass--energy relation $\mu_{s}=\omega$ turns the superradiance condition into the bound $\tilde{q}_{s}/\mu_{s}>r_{\min}/\tilde{Q}$, which equals unity for Reissner--Nordstr\"om and is pushed below unity once the non-Abelian charge is switched on, so the model admits lighter charged states than the Maxwell case while still protecting the horizon.

Read as a whole, these results say that the \WCCC\ survives scalar absorption even with a non-Abelian power-law gauge field present and the central charge free to vary. The \WGC\ does not trigger naked-singularity formation here. It acts instead as a selector, fixing the direction of evolution toward or away from extremality. The censorship boundary and the charge-to-mass threshold organize into a single phase diagram over the $(\tilde{Q},\gamma)$ plane, and the absorption flux moves a configuration deeper into the black-hole region rather than across the boundary into the naked one. The exponent $\gamma$ and the central charge $C$ enter the bookkeeping only through the sensitivities $\partial f/\partial\gamma$ and $\partial f/\partial C$. Both are fixed couplings that do not vary spontaneously during absorption, so the conclusion is stable against the large excursions that could in principle endanger it. The same thermodynamics carries a clean local-stability story: the heat capacity at fixed charges develops a Davies divergence near $S\simeq1$ that drifts upward with $\gamma$, separating a small-entropy stable branch from a large-entropy unstable one, while the Yang--Mills charge lowers the extremal radius and so shifts the photon sphere that sets the optical appearance of the black hole.

Several tasks suggest themselves for the near future. The most direct is to lift the analysis to rotating and higher-dimensional \EMPYM\ backgrounds, where frame dragging and the larger sphere of horizons may sharpen or soften the threshold found here. A second line is to replace the power-law Yang--Mills sector with more general nonlinear electrodynamics and to ask whether the relaxation of the charge-to-mass bound persists, since that would tell us how much of the effect is specific to the exponent $\gamma$. It would also be worth following the phase structure under quantum-gravity corrections to the entropy, which could move the Davies point and reshape the stable branch. Finally, the lowered threshold of Fig.~\ref{fig:wgc} and the accompanying shift of the photon sphere invite a quantitative confrontation with horizon-scale imaging and strong lensing, so that the interplay between cosmic censorship and the weak gravity conjecture in this class of black holes can eventually be tested against data rather than left as a purely theoretical statement.

Alongside the censorship and weak-gravity analysis, we read the information content of the same
EMPYM--AdS black holes through the island prescription. The entanglement entropy of the Hawking
radiation traces a standard Page curve: linear growth first, then saturation once the island
forms past the Page time. The useful output is an explicit analytic Page time, written in terms
of the full set of thermodynamic variables, the Maxwell charge, the Yang--Mills parameter, the
nonlinear exponent $\gamma$, and the pressure. It ties black hole chemistry to information
recovery directly. The same data that governs censorship and stability also sets the pace of the
unitarity-restoring process, and for large Maxwell charge or small $\gamma$ it pushes the hole
toward a cold, near-extremal limit where the Page time diverges and recovery stalls.

\section*{Acknowledgments}
S.N.G.\ and B.P.\ acknowledge the support of Damghan University; B.P.\ further acknowledges the Center for Theoretical Physics at Khazar University. \.I.S.\ is grateful to T\"UB\.ITAK and the Eastern Mediterranean University for support, and acknowledges the networking support of COST Actions CA22113 (``Fundamental challenges in theoretical physics''), CA21106 (``COSMIC WISPers in the Dark Universe''), CA23130 (``Bridging high and low energies in search of quantum gravity (BridgeQG)''), CA21136 (``Addressing observational tensions in cosmology with systematics and fundamental physics (CosmoVerse)''), and CA23115 (``Relativistic Quantum Information (RQI-Action)'').

\section*{Data Availability Statement}
The computational scripts that reproduce every analytic result and figure reported in this article are available from the corresponding author upon reasonable request.

{\footnotesize
\setlength{\bibsep}{1.2pt plus 0.3ex}
\begin{multicols}{2}
\bibliographystyle{unsrtnat}
\bibliography{finalref}
\end{multicols}
}

\end{document}